\documentclass[twocolumn,english,showpacs,preprintnumbers,amsmath,amssymb,floatfix,nofootinbib]{revtex4-1}

\usepackage[T1]{fontenc}
\usepackage[latin9]{inputenc}
\usepackage{color}
\usepackage{array}
\usepackage{amstext}
\usepackage{graphicx}
\usepackage{esint}
\usepackage{rotating}
\usepackage[font=small,labelfont=bf]{caption}
\usepackage{xcolor}
\usepackage[colorlinks = true,
linkcolor = magenta,
urlcolor  = blue,
citecolor = red,
anchorcolor = blue]{hyperref}

\makeatletter

\@ifundefined{textcolor}{}
{%
 \definecolor{BLACK}{gray}{0}
 \definecolor{WHITE}{gray}{1}
 \definecolor{RED}{rgb}{1,0,0}
 \definecolor{GREEN}{rgb}{0,1,0}
 \definecolor{BLUE}{rgb}{0,0,1}
 \definecolor{CYAN}{cmyk}{1,0,0,0}
 \definecolor{MAGENTA}{cmyk}{0,1,0,0}
 \definecolor{YELLOW}{cmyk}{0,0,1,0}
 }

\@ifundefined{definecolor}
 {\usepackage{color}}{}
\@ifundefined{definecolor}
 {\usepackage{color}}{}
\makeatother
\usepackage{babel}

\def\Re{{\cal R \mskip-4mu \lower.1ex \hbox{\it e}\,}}
\def\Im{{\cal I \mskip-5mu \lower.1ex \hbox{\it m}\,}}

\def\tev{\,{\ifmmode\mathrm {TeV}\else TeV\fi}}
\def\gev{\,{\ifmmode\mathrm {GeV}\else GeV\fi}}
\def\mev{\,{\ifmmode\mathrm {MeV}\else MeV\fi}}
\def\to{\rightarrow}

\begin{document}


\title { Constraining Higgs boson effective couplings at electron-positron colliders }

\author { Hamzeh Khanpour$^{1,2}$ }
\email{Hamzeh.Khanpour@mail.ipm.ir}

\author{ Mojtaba Mohammadi Najafabadi$^{2}$ }
\email{Mojtaba@cern.ch}

\affiliation {
$^{(1)}$Department of Physics, University of Science and Technology of Mazandaran, P.O.Box 48518-78195, Behshahr, Iran        \\
$^{(2)}$School of Particles and Accelerators, Institute for Research in Fundamental Sciences (IPM), P.O.Box 19395-5531, Tehran, Iran }

\date{\today}

%
\begin{abstract}\label{abstract}

We probe the dimension-six operators contributing to Higgs production in association with a $Z$ boson at the future high-luminosity electron-positron colliders.
Potential constraints on dimension-six operators in the Higgs sector are determined by performing a shape analysis on the differential angular distribution
of the Higgs and $Z$ boson decay products. The analysis is performed at the center-of-mass energies of 350 and 500 GeV including a realistic detector simulation and the main sources of background processes. The 68\% and 95\% confidence level upper limits are obtained on the contributing anomalous couplings considering only the decay of the Higgs boson into a pair of $b$-quarks and leptonic $Z$ boson decay. Our results show that angular observables provide a great sensitivity to the anomalous couplings, in particular, at the high-luminosity regime.

\end{abstract}
%


\maketitle

\tableofcontents{}

%
\section{Introduction}\label{sec:intro}

After the Higgs boson discovery at the Large Hadron Collider (LHC) run-I in 2012~\cite{Aad:2012tfa,Chatrchyan:2012xdj}, the main task is to provide precise 
measurement of its couplings to the Standard Model (SM) particles as well as its other properties.  
This opens a way to look for potential new physics effects and  provides the 
possibility for revealing effects which may show up at high energy scales.
The recent results of the ATLAS and CMS experiments  in probing  the couplings of Higgs boson shows 
no signs of new physics~\cite{ATLAS-CONF-2015-044}. The Higgs couplings to the SM particles also have been studied extensively in 
several analyses using available data from the LHC and previous experiments~\cite{Hagiwara:1993qt,Han:2004az,Corbett:2012ja,Dumont:2013wma,Ciuchini:2014dea,Ellis:2014dva,Falkowski:2014tna,Falkowski:2015fla,Brehmer:2015rna,Falkowski:2015wza,Achard:2003ty,Achard:2004cf}.

The compatibility of the current measurements with the SM predictions in the 
Higgs sector causes the new physics scale to be different from the electroweak
scale. This suggests to search for  new physics effects beyond the SM by adopting the
effective field theory approach  without going through the details of any specific scenarios.
In this approach, the effective operators consist of only the SM fields and are
obtained by integrating out heavy degrees of freedom. These effective interactions are suppressed by inverse powers of  the new physics scale. 
Such an effective Lagrangian is required to respect to the Lorentz symmetry and the ${\rm SU(3)}_{\rm C} \times {\rm SU(2)}_{\rm L} \times {\rm U(1)}_{\rm Y}$ SM gauge  symmetries.
Assuming baryon and lepton number conservation, operators of dimension six are the first corrections which are added to the 
SM action. The effective Lagrangian can be written as follows:

\begin{eqnarray}
	\mathcal{L}_{eff} = \mathcal{L}_{SM}+ \sum_{i} \frac{c_{i} \mathcal{O}_{i}}{\Lambda^{2}} \,,
\end{eqnarray}

where the effects of possible new physics is assumed to appear at an energy scale
of $\Lambda$ , $c_{i}$ coefficients are 
dimensionless Wilson coefficients, and $\mathcal{O}_{i}$  are dimension six operators obtained by integrating out the heavy  
degrees of freedom  in the underlying theory.

So far, there are many studies to constrain  these Wilson coefficients  in the Higgs boson sector from the LHC run I data and 
from the electroweak precision tests at large electron-positron (LEP) and future colliders~\cite{Han:2004az,Corbett:2012ja,Dumont:2013wma,Pomarol:2013zra,Alam:1997nk,     
Monfared:2016vwr,Ferreira:2016jea, Khatibi:2014bsa,Hesari:2015oya,Liu:2016dag,Khanpour:2014xla,Elias-Miro:2013gya,Elias-Miro:2013eta,Chen:2013kfa,Mebane:2013zga,Craig:2014una,Gounaris:2014tha,Kilian:1995tr,Beneke:2014sba,Heinemeyer:2001iy,Dawson:2002wc,Mimasu:2015nqa,Greljo:2015sla,Cohen:2016bsd,Ge:2016zro,Ge:2016tmm,Craig:2015wwr,Hesari:2014eua,Arbey:2016kqi}.
If the LHC at run II does not observe any significant deviation from the SM
expectations, stronger bounds on the coefficients of the effective operators would be set.
Realistic estimations of constraints on the effective coefficients of Higgs related operators
after the LHC run II with high integrated luminosity have been provided in \cite{Englert:2015hrx}.

Electron-positron colliders such as Compact Linear Collider  (CLIC)~\cite{Linssen:2012hp,Aicheler:2012bya,Abramowicz:2013tzc},  International Linear Collider (ILC)~\cite{Fujii:2015jha,Behnke:2013lya,Barklow:2015tja,Asner:2013psa,Moortgat-Picka:2015yla}, Circular Electron-Positron Collider (CEPC)~\cite{CEPC-SPPCStudyGroup:2015csa,CEPC-SPPCStudyGroup:2015esa} or 
high-luminosity high-precision FCCee~\cite{Gomez-Ceballos:2013zzn,Koratzinos:2015ywz,d'Enterria:2016yqx,Ellis:2015sca,Janot:2015yza,d'Enterria:2015toz,Benedikt:2015iia,Koratzinos:2015fya,Zimmermann:2015mea,d'Enterria:2016cpw},  
with clean experimental environment due to the absence of hadronic initial state and  accurately
known collision energy provide a good opportunity to probe precisely the Higgs boson couplings as well as measurement of the SM parameters
with high accuracy.  Going up to  high energies and luminosities, these colliders can continue the studies made by LEP and 
provide an excellent place in  search for new physics beyond the SM~\cite{Brooijmans:2016vro,Asner:2013psa,Han:2015ofa,Moortgat-Picka:2015yla,Fujii:2015jha,Barklow:2015tja,Gomez-Ceballos:2013zzn,Baak:2013fwa,Fan:2014vta,Fan:2014axa,Hartmann:2016pil,Banerjee:2015bla,Amar:2014fpa}.

In this work, by adopting effective Lagrangian approach in the strongly interacting light Higgs (SILH) basis~\cite{Alloul:2013naa,Artoisenet:2013puc} \footnote{This basis is not unique and could be connected to other bases.}, we constrain coefficients of dimension six operators using the Higgs production in association 
with a $Z$ boson at the electron-positron colliders with the center-of-mass energies of 350 GeV and 500 GeV.
In Higgs production in association with a $Z$ boson, the correction coming from dimension six operators
are scaled as $s/\Lambda^{2}$ where $s$ is the center-of-mass energy of the collisions and must be 
greater than $(m_{Z}+m_{H})^{2}$ to produce $H+Z$ on-shell.

The results are obtained using a realistic simulation  including the main background contributions for the $e^+ e^- \rightarrow  H+Z$ process. 
The analysis is based on the channel in which the Higgs boson decays into a pair of b-quarks and $Z$ boson decays leptonically. The upper limits on the 
coefficients of dimension six operators are obtained at 68\% and 95\% confidence level using a $\chi^{2}$ analysis on the
angular distribution of the Higgs and $Z$ bosons decay products.
The results are presented for the integrated luminosities of 300 fb$^{-1}$ and 3 ab$^{-1}$.     

The present paper is organized as follows: In Section~\ref{sec:framework}, a brief description of the theoretical framework and assumptions are given. 
Details of event generation, detector simulation, event selection and the strategy of the analysis are illustrated in Section~\ref{sec:analysis}. 
The statistical method used to obtain upper limits on the coefficients of dimension-six operators is presented in Section~\ref{sec:statistical}.
Our results for integrated luminosities of 300 fb$^{-1}$ and 3 ab$^{-1}$ are discussed in Section~\ref{sec:results}. Finally, summary and conclusions are given in Section~\ref{sec:Discussion}.

%
\section{Theoretical framework and assumptions}\label{sec:framework}
%

In this section, the most general effective Lagrangian up to dimension-six containing
the SM fields, which respects the gauge and global symmetries of the SM, is introduced.
There are equivalent ways to write this effective Lagrangian which cause to have different bases.
In this work, the convention for EFT operators proposed in Refs.~\cite{Alloul:2013naa,Artoisenet:2013puc,Englert:2015hrx} is followed. Considering baryon and lepton number conservation,
the relevant parts of the effective Lagrangian  which affect the Higgs boson couplings have the following terms:

\begin{eqnarray}\label{eq:L}  
	\mathcal{L}_{\rm EFT} = \mathcal{L}_{\rm SM} + \mathcal{L}_{\rm SILH} + \mathcal{L}_{\rm F_{1}} + \mathcal{L}_{\rm F_{2}},
\end{eqnarray}   

where $ \mathcal{L}_{\rm SILH}$ consists of a set of CP-even dimension six operators involving the Higgs doublet and is inspired
from models in which the Higgs field is part of a strongly interacting sector \cite{Giudice:2007fh}.  Third term, $\mathcal{L}_{\rm F_{1}}$,  contains interactions
among two Higgs fields and a pair of leptons or quarks. The fourth term of the effective Lagrangian, $\mathcal{L}_{\rm F_{2}}$, expresses
the interactions of a quark or lepton pair with a single Higgs field and a gauge boson. 
For instance,  $\mathcal{L}_{\rm SILH}$ has the following form:

\begin{eqnarray}\label{leff}
	\begin{split}
		\mathcal{L}_{\rm SILH} = & \
		\frac{g_s^2\ \bar c_{g}}{m_{W}^2} \Phi^\dag \Phi G_{\mu\nu}^a G_a^{\mu\nu}
		+\frac{g'^2\ \bar c_{\gamma}}{m_{W}^2} \Phi^\dag \Phi B_{\mu\nu} B^{\mu\nu}     \\     \label{eq:eft}
		& \ 
		+ \frac{i g'\ \bar c_{B}}{2 m_{W}^2} \big[\Phi^\dag \overleftrightarrow{D}^\mu \Phi \big] \partial^\nu  B_{\mu \nu}  \\
		& \
		+ \frac{i g\ \bar  c_{W}}{2m_{W}^2} \big[ \Phi^\dag \sigma_{k} \overleftrightarrow{D}^\mu \Phi \big]  D^\nu  W_{\mu \nu}^k    \\
		& \
		+ \frac{ i g\ \bar c_{HW}}{m_{W}^2} \big[D^\mu \Phi^\dag \sigma_{k} D^\nu \Phi\big] W_{\mu \nu}^k    \\
		& \
		+ \frac{i g'\ \bar c_{HB}}{m_{W}^2}  \big[D^\mu \Phi^\dag D^\nu \Phi\big] B_{\mu \nu}    \\
		& \
		+ \frac{\bar c_{H}}{2 v^2} \partial^\mu\big[\Phi^\dag \Phi\big] \partial_\mu \big[ \Phi^\dagger \Phi \big]
		+ \frac{\bar c_{T}}{2 v^2} \big[ \Phi^\dag {\overleftrightarrow{D}}^\mu \Phi \big] \big[ \Phi^\dag {\overleftrightarrow{D}}_\mu \Phi \big]    \\
		& \
		- \frac{\bar c_{6} \lambda}{v^2} \big[\Phi^\dag \Phi \big]^3      \\  
		& \  - \bigg[
		\frac{\bar c_{l}}{v^2} y_\ell\ \Phi^\dag \Phi\ \Phi {\bar L}_L e_R
		+\frac{\bar c_{u}}{v^2} y_u \Phi^\dag \Phi\ \Phi^\dag\cdot{\bar Q}_L u_R    \\
		& \
		+ \frac{\bar c_{d}}{v^2} y_d \Phi^\dag \Phi\ \Phi {\bar Q}_L d_R
		+ {\rm h.c.} \bigg]  \,,
	\end{split}
\end{eqnarray}

where $\Phi$ is a weak doublet which contains the Higgs boson field and  $B^{\mu\nu}$,  $W^{\mu \nu}$, $G^{\mu\nu}$ are the 
electroweak and strong field strength tensors. The hermitian covariant derivative is defined as 
$\Phi^\dag \overleftrightarrow{D}^\mu \Phi = \Phi^\dag (D^{\mu}\Phi) - (D^{\mu}\Phi)^{\dag}\Phi$.
The Higgs quartic coupling is denoted by $\lambda$ and $v$ is the weak scale which is defined as $v = 1/(\sqrt{2}G_{F})^{1/2} = 246$ GeV.
$\bar{c}_{u}$, $\bar{c}_{d}$ and $\bar{c}_{l}$ are real parameters as the Higgs boson is assumed to be a CP-even particle.

Accuracy of the oblique parameters $S$ and $T$ from the electroweak precision measurements
leads to reduce the number of parameters in the above effective Lagrangian. The per-mille constraints on $S$ and $T$ parameters 
lead  $\bar c_{T} = 0$ and $\bar c_{B} + \bar c_{W} = 0$ as these are directly related to the oblique parameters~\cite{Giudice:2007fh,
Contino:2013kra,Ellis:2014jta,Barbieri:2004qk}. 

The effective Lagrangian describing the Higgs boson couplings has been  studied 
at CLIC with 1 ab$^{-1}$ of integrated luminosity at the center-of-mass energy of 3 TeV \cite{Moortgat-Picka:2015yla}.
The study has been performed through double Higgs production as the vertices involving more than a Higgs boson
can provide the possibility for testing the composite nature of the Higgs boson.
The sensitivity reach has been reported in the plane of $\xi$ and $m_{\rho}$ where 
$m_{\rho}$ is the mass scale of the heavy strong sector resonances and $\xi = \frac{v}{f}$. $f$ is the compositeness scale and 
$v$ is the vacuum expectation value.
A detailed description of these parameters could be found in \cite{Giudice:2007fh}. According to this study, the region
of $\xi > 0.03$ could be excluded at $95\%$ confidence level (CL) for any value of $m_{\rho}$.

\begin{figure*}[htb]
	\begin{center}
		\vspace{0.40cm}
		\resizebox{0.60\textwidth}{!}{\includegraphics{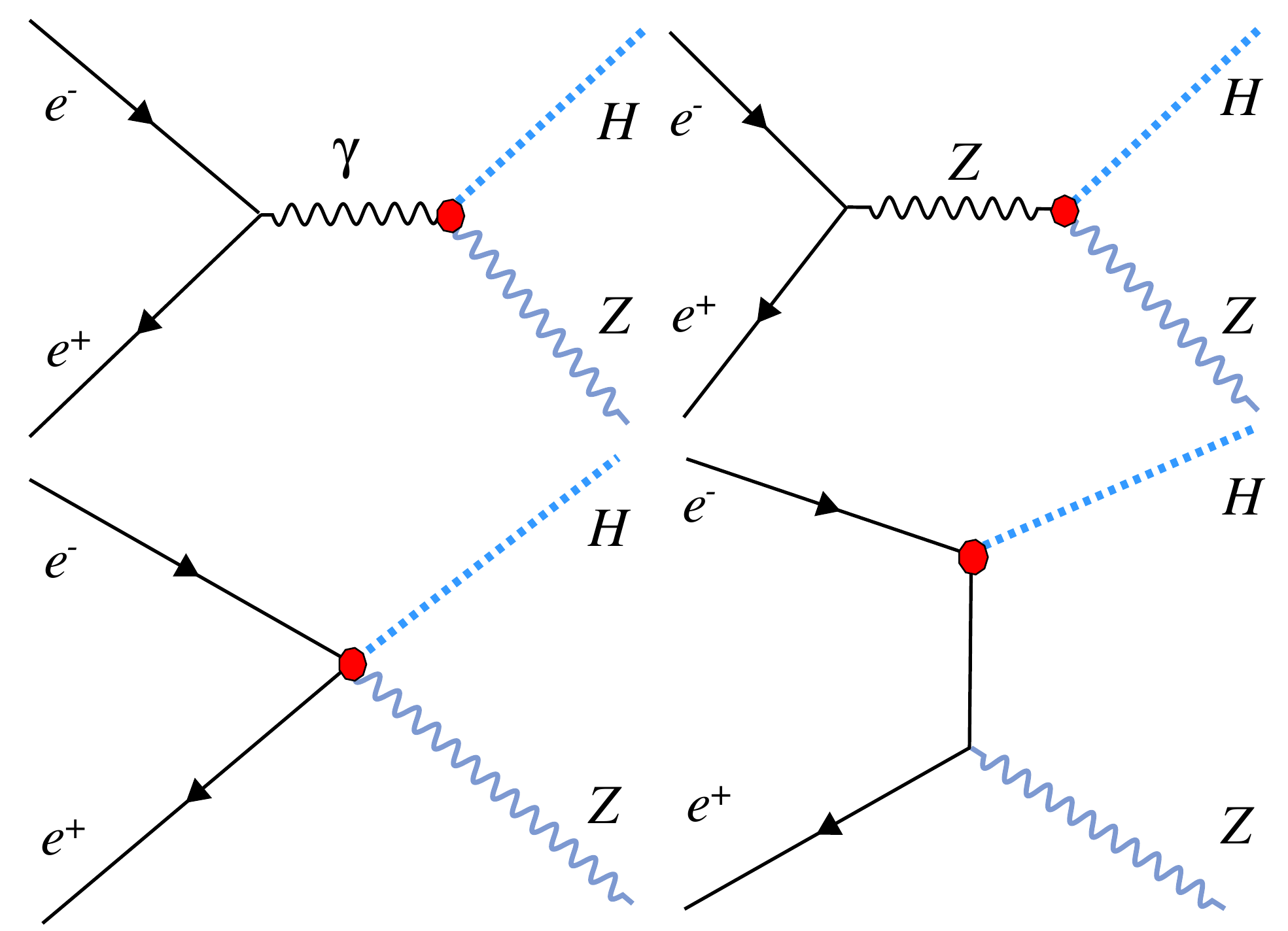}}   		
		\caption{ Representative tree level Feynman diagrams for the production of a Higgs boson in association with
			a $Z$ boson at an electron-positron collider in the presence of dimension six operators.} \label{feynman}
	\end{center}
\end{figure*}

In this analysis, we consider the effects of $\mathcal{L}_{\rm EFT}$ (Eq.\ref{eq:L}) in the 
$e^{-}+e^{+} \rightarrow H+Z$ process and the contributions from any other possible effective operators are neglected.
The SM tree level part contribution is not dependent on the momenta of the particles, while  $\mathcal{L}_{\rm EFT}$
introduces  momentum-dependent interactions. As a result, the new contributions from $\mathcal{L}_{\rm EFT}$ 
affect the decay rates, production cross sections as well as the shape of differential distributions. 
In this paper, by exploiting differences in the shape of angular distributions of the decay products of the
Higgs and $Z$ bosons,
the  new involved couplings from $\mathcal{L}_{\rm EFT}$ in $e^{-}+e^{+} \rightarrow H+Z$ process are studied.
The representative Feynman diagrams for production of a Higgs boson in association with a $Z$ boson are
depicted in Fig.~\ref{feynman}. The vertices affected by $\mathcal{L}_{\rm EFT}$ are presented by 
filled circles.

The $e^{-}+e^{+} \rightarrow H+Z$ process is sensitive to the following set of $\mathcal{L}_{\rm EFT}$ parameters:
\begin{eqnarray}
	\bar{c}_{\gamma}, \bar{c}_{HW}, \bar{c}_{HB}, \bar{c}_{W},\bar{c}_{B},\bar{c}_{H}, \bar{c}_{T},\bar{c}_{eW},\bar{c}_{eB},\bar{c}_{l} \,.
\end{eqnarray}
The parameters $\bar{c}_{eW}$ and $\bar{c}_{eB}$ are coming from $\mathcal{L}_{\rm F_{2}}$ in Eq.~(\ref{eq:L})
where the related terms contain electron Yukawa coupling $y_{e}$. 
As we mentioned before, the precise measurement of oblique parameters $S$ and $T$ leads
$ \bar{c}_{T}=0$ and $\bar{c}_{W}= -\bar{c}_{B}$ which reduces the number of degrees of freedom 
from ten to eight. Because of very small Yukawa coupling of electron, $\bar{c}_{l}$, $\bar{c}_{eW}$ and $\bar{c}_{eB}$
do not lead to considerable modifications in the cross section. 
Consequently, we limit ourselves to only the remaining five parameters: 
$\bar{c}_{\gamma}, \bar{c}_{HW}, \bar{c}_{HB}, \bar{c}_{W},\bar{c}_{H}$. 

Another approach to present the effective Lagrangian which is interesting from the phenomenological and experimentally 
is the effective Lagrangian in the mass basis.  This approach has been found to be a useful approach for
electroweak precision tests. Following Ref.~\cite{Alloul:2013naa}, the relevant subset of the
anomalous Higgs  boson couplings in the mass basis includes:
\begin{eqnarray}
\mathcal{L} &=&-\frac{1}{4}g^{(1)}_{hzz}Z_{\mu\nu}Z^{\mu\nu}h-g^{(2)}_{hzz}Z_{\nu}\partial_{\mu}Z^{\mu\nu}h
+\frac{1}{2}g^{(3)}_{hzz}Z_{\mu}Z^{\mu}h \nonumber \\
&-&\frac{1}{2}g^{(1)}_{haz}Z_{\mu\nu}F^{\mu\nu}h-g^{(2)}_{haz}Z_{\nu}\partial_{\mu}F^{\mu\nu}h,
\end{eqnarray}
where the relation between the couplings in the mass basis and the dimension-six coefficients are 
given as below:
\begin{eqnarray}
g^{(1)}_{hzz} &=& \frac{2g}{c_{W}^{2}m_{W}}[\bar{c}_{HB}s_{W}^{2}-4\bar{c}_{\gamma}s_{W}^{4}
+c_{W}^{2}\bar{c}_{HW}], \nonumber \\
g^{(2)}_{hzz} &=& \frac{g}{c_{W}^{2}m_{W}}[(\bar{c}_{HW}+\bar{c}_{W})c_{W}^{2}
+(\bar{c}_{B}+\bar{c}_{HB})s_{W}^{2})], \nonumber \\
g^{(3)}_{hzz} &=& \frac{gm_{W}}{c_{W}^{2}}[1-\frac{1}{2}\bar{c}_{H}-2\bar{c}_{T}+
8\bar{c}_{\gamma}\frac{s_{W}^{4}}{c^{2}_{W}}], \nonumber \\
g^{(1)}_{haz} &=& \frac{gs_{W}}{c_{W}m_{W}}[\bar{c}_{HW}-\bar{c}_{HB}+
8\bar{c}_{\gamma}s_{W}^{2}], \nonumber \\
g^{(2)}_{haz} &=& \frac{gs_{W}}{c_{W}m_{W}}[\bar{c}_{HW}-\bar{c}_{HB}-
\bar{c}_{B}+\bar{c}_{W}],
\end{eqnarray}
Detailed information together with  a complete list of anomalous
couplings of Higgs boson in the mass basis could be found in \cite{Alloul:2013naa}.

We calculate the effects of the dimension six operators on $H+Z$  production with
Monte Carlo (MC) simulations using { \tt MadGraph5-aMC@NLO}~\cite{Alwall:2014bza,Alwall:2014hca,Alwall:2011uj}.
The Lagrangian introduced in Eq.~\ref{eq:L} has been implemented in {\tt FeynRule} package \cite{Alloul:2013bka} and then to { \tt MadGraph5-aMC@NLO}
which can be found in Refs.~\cite{Alloul:2013naa,Artoisenet:2013puc}.
In the next sections, the details of simulation and determination of the 68\% and 95\% confidence level (CL)
limits on the coefficients of dimension six operators are described.

%
\section{Simulation Details and Analysis}\label{sec:analysis}
%

In this section, the details of simulation for probing the effective Lagrangian through
the $H+Z$ events in the electron-positron collisions are discussed.
We focus on the Higgs decay into a pair of b-quarks and $Z$ boson decay into a pair
of charged leptons,($\ell = e, \ \mu$). As a result, 
the final state consists of two energetic jets originating from the hadronization of two b-quarks as well as two 
charged leptons.   The dominant background processes which are considered in this analysis are: 
(i) $e^+ e^- \to ZZ$ in which one $Z$ decays hadronically and another one decays into charged leptons; 
(ii) $e^+ e^- \to t \bar{t}$ in dilepton final state which contains two b-jets, two charged lepton and missing energy;
(iii) $e^+ e^- \to Z\gamma  \rightarrow \ell^{+}\ell^{-}jj$ and $e^+ e^- \to \gamma \gamma  \rightarrow \ell^{+}\ell^{-}jj$;
(iiii) $e^+ e^- \to W^{+} W^{-}Z$ either with leptonic decay of both $W$ bosons and hadronic decay of $Z$ boson or with hadronic decay of the 
$W$ bosons and leptonic decay of $Z$ boson. 

Signal and background processes are generated with { \tt MadGraph5-aMC@NLO}
~\cite{Alwall:2014bza,Alwall:2014hca,Alwall:2011uj} event generator
and are passed through {\tt PYTHIA 8}~\cite{Sjostrand:2007gs,Sjostrand:2003wg} for
parton showering, hadronization, and decay of unstable particles. {\tt Delphes 3.3.2}
~\cite{deFavereau:2013fsa,Mertens:2015kba} is employed 
to account for the detector effects similar to an ILD-like detector~\cite{Behnke:2013lya}. 
The SM input parameters are taken as the following \cite{Olive:2016xmw}:
$m_H = 125.0 \, {\rm GeV}, \, m_t = 173.34 \, {\rm GeV}, \, m_W = 80.385 \, {\rm GeV}$
and $m_Z = 91.187 \, {\rm GeV}$.

The tracking efficiency of an ILD-like detector is set to 99\% for charged particles with $p_T > 0.1$ GeV and $|\eta| \leq 2.4$, 
including electrons and muons. Electrons, muons, and photons with transverse momenta
greater that 10 GeV are reconstructed with an efficiency of $99\%$ in an ILD-like detector.
The momentum resolution for muons are: 
$\Delta p /p = (1.0+ 0.01\times p_T[\rm GeV])\times 10^{-3}$ for  $|\eta| \leq 1$  and
$\Delta p /p = (1.0+ 0.01\times p_T[\rm GeV])\times 10^{-2}$ for  $1 < |\eta| \leq 2.4$.
The electron and jets energy resolutions are assumed to be:

\begin{eqnarray}
	\frac{\Delta E_{\rm electron}}{E_{\rm electron}} & = & \frac{15\%}{\sqrt{E_{\rm electron} (\rm GeV)}} + 1.0\% \nonumber \\ 
	\frac{\Delta E_{\rm jets}}{E_{\rm jets}} & = & \frac{50\%}{\sqrt{E_{\rm jets} (\rm GeV)}} + 1.5\% \,,
\end{eqnarray}

Jets are reconstructed with the anti-$k_{t}$ algorithm~\cite{Cacciari:2008gp} 
using the {\tt FastJet}  package~\cite{Cacciari:2011ma} with a cone size parameter R = 0.5. 
The b-tagging efficiency and misidentification rates depend on the jet transverse momentum 
and are taken according to an ILD-like detector~\cite{Behnke:2013lya}. At a transverse momentum of around 50 GeV,
the b-tagging efficiency is around $64\%$, c-jet misidentification rate is $17\%$, and 
a misidentification rate of light-jet is around $1.2\%$.

We select the signal and background events according to the following requirements:
Exactly two same flavor opposite sign charged leptons ($\ell = e, \mu$) with the transverse momentum 
$p_T^\ell > 10$ GeV and the pseudo-rapidity of $|\eta_{\ell}| \leq 2.5$ are required. 
Each event is required to have only two b-tagged jets with 
$p_T^{\rm jets} > 20$ GeV and  $|\eta_{\rm jets}| \leq 2.5$.  To make sure all
objects are well-isolated, we require that the angular separation 
$\Delta R_{\ell, {\rm b-jets}} = \sqrt{(\Delta \phi)^2 + (\Delta \eta)^2} > 0.5$. 
The above cuts are denoted as the  preselection cuts.

\begin{figure*}[htb]
	\begin{center}
		\vspace{0.50cm}
		\resizebox{0.450\textwidth}{!}{\includegraphics{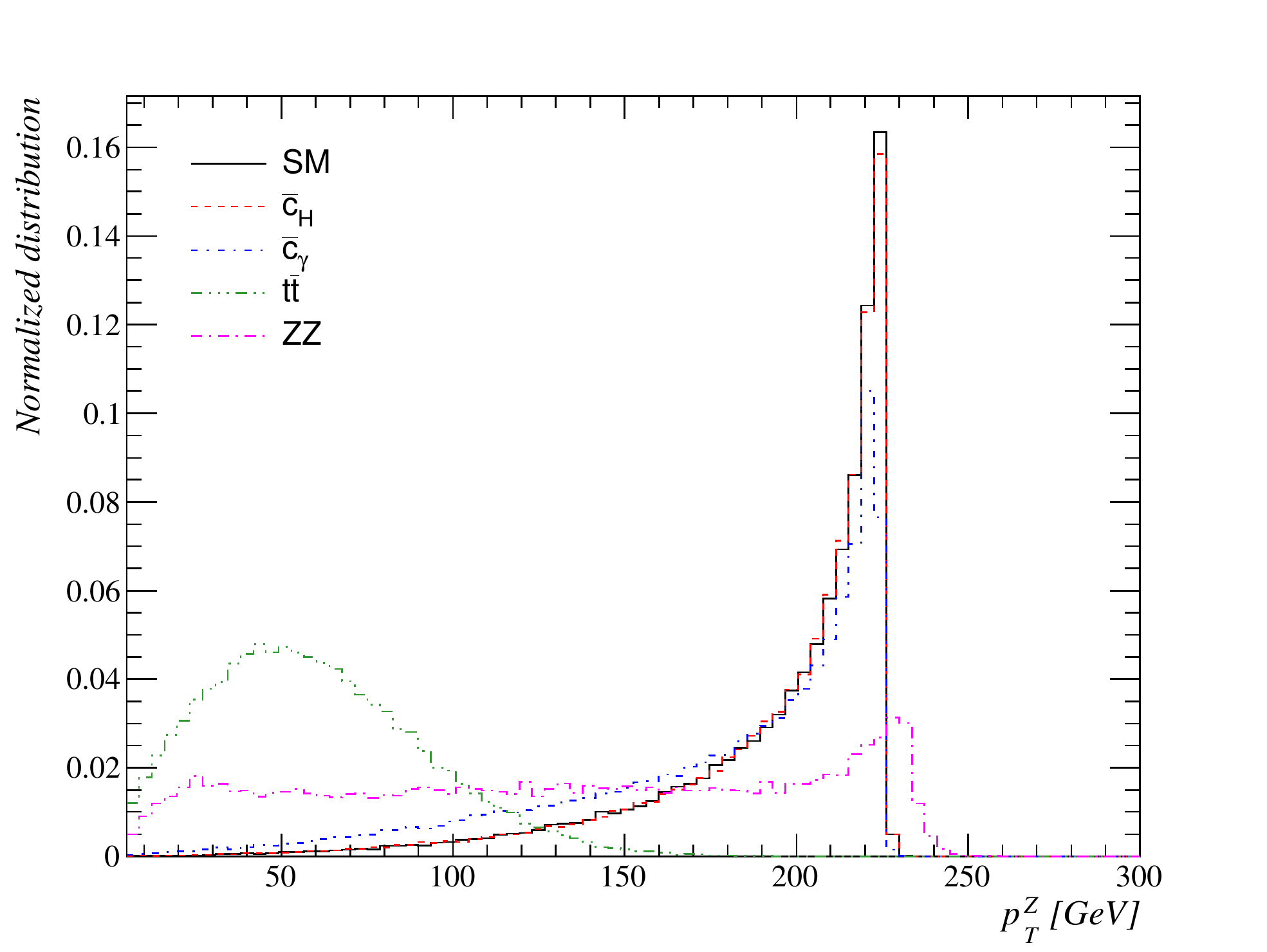}}   
		\resizebox{0.450\textwidth}{!}{\includegraphics{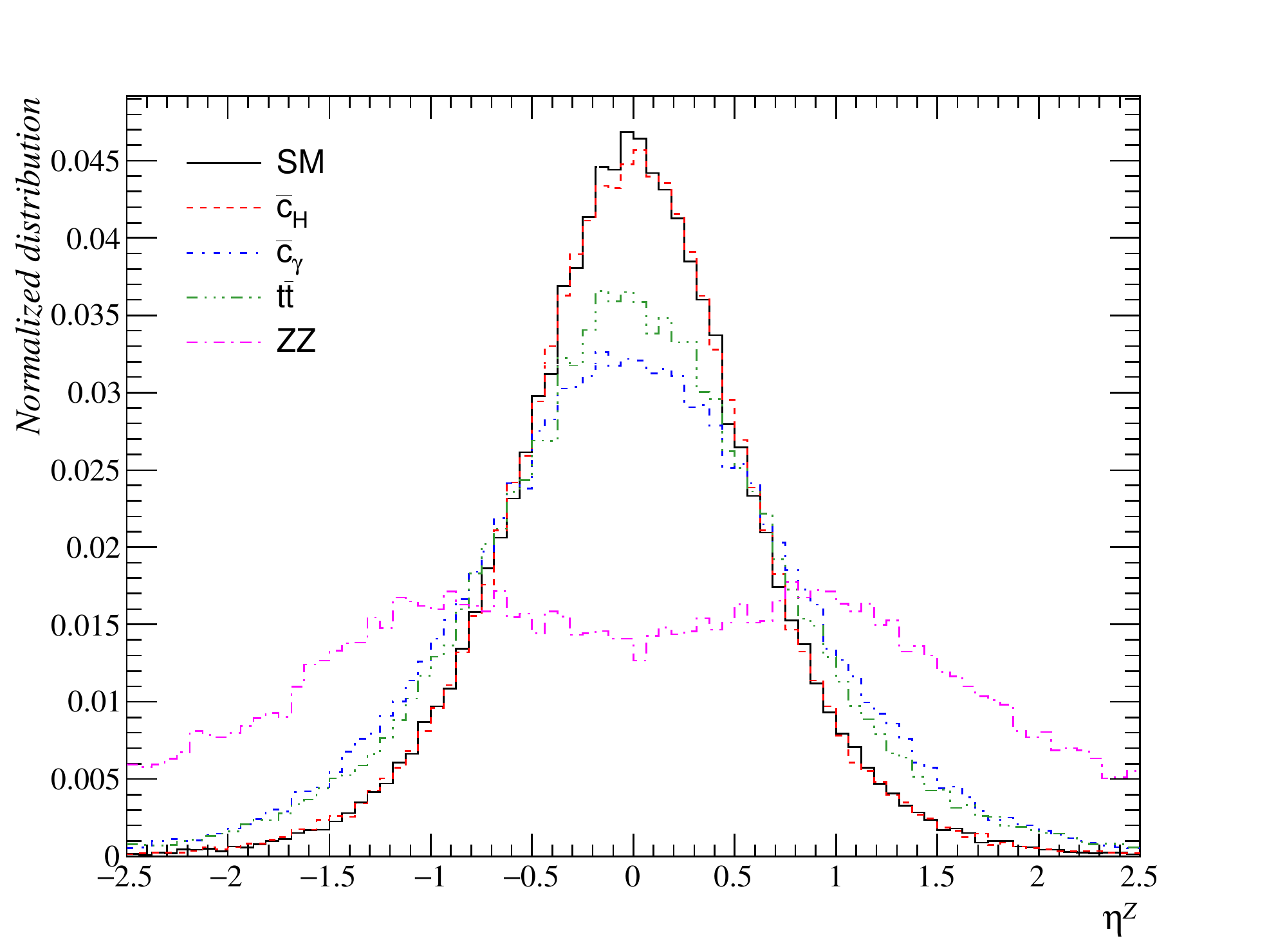}}   
		\resizebox{0.450\textwidth}{!}{\includegraphics{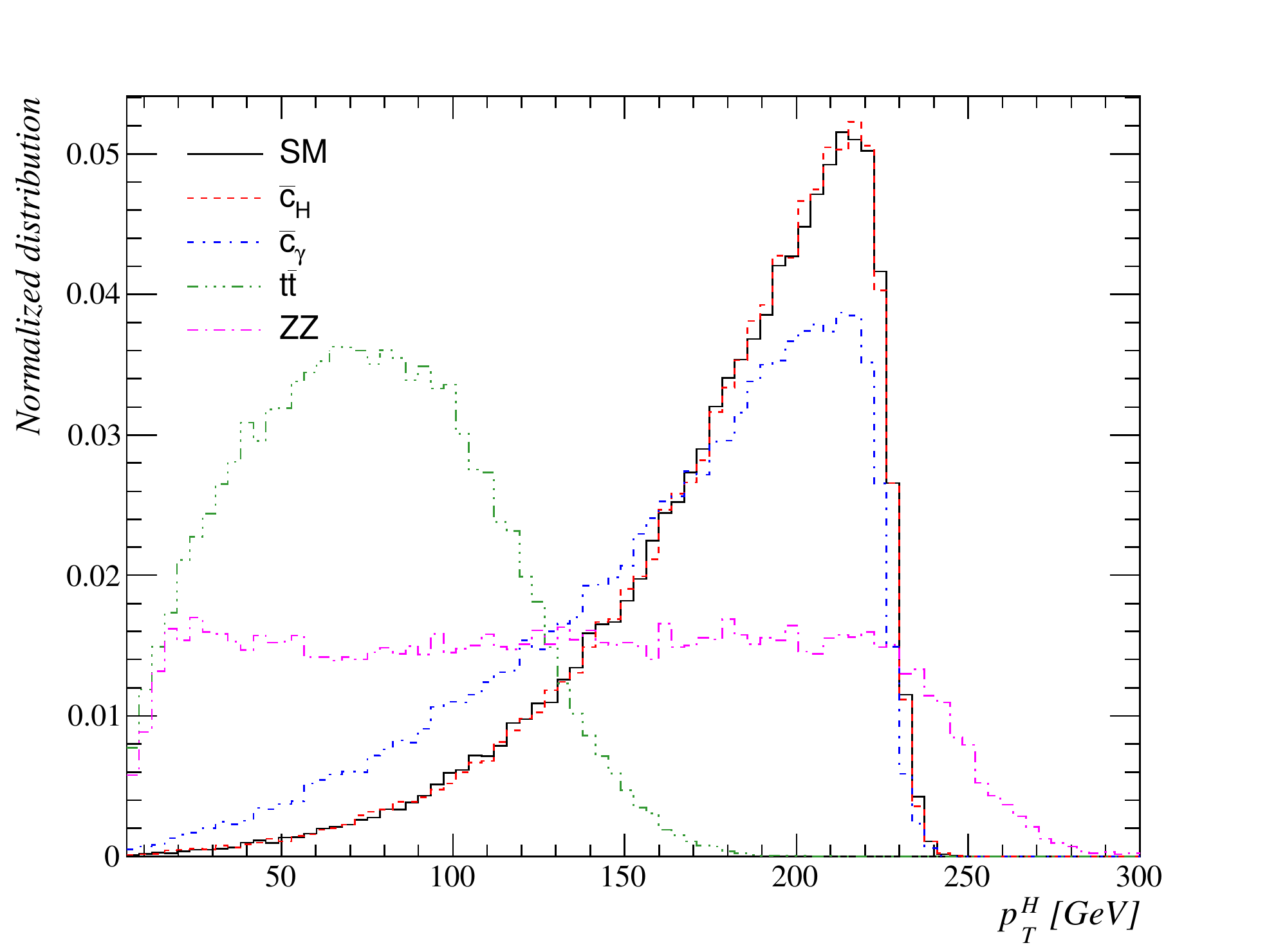}}   
		\resizebox{0.450\textwidth}{!}{\includegraphics{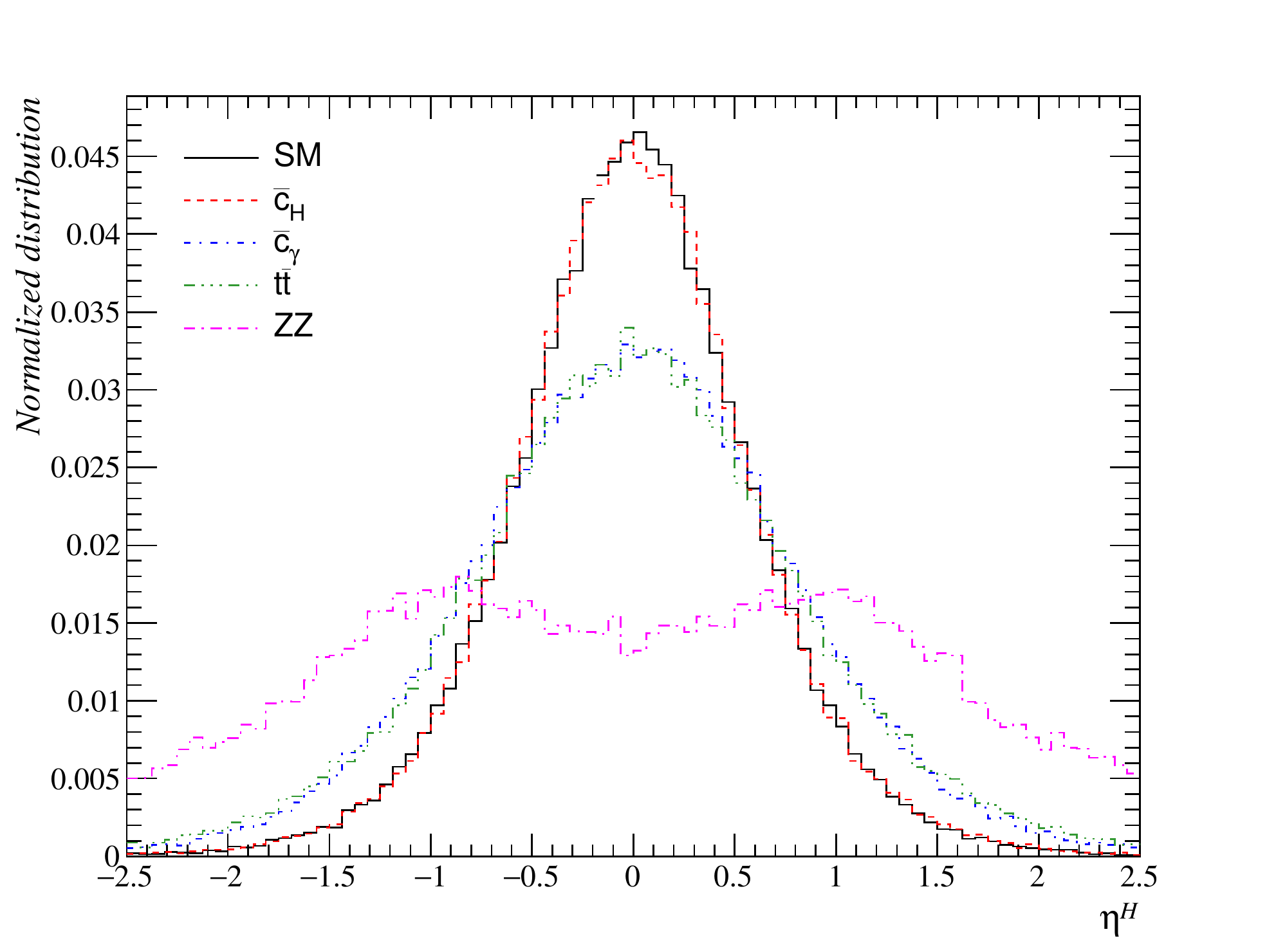}}   
	      \resizebox{0.450\textwidth}{!}{\includegraphics{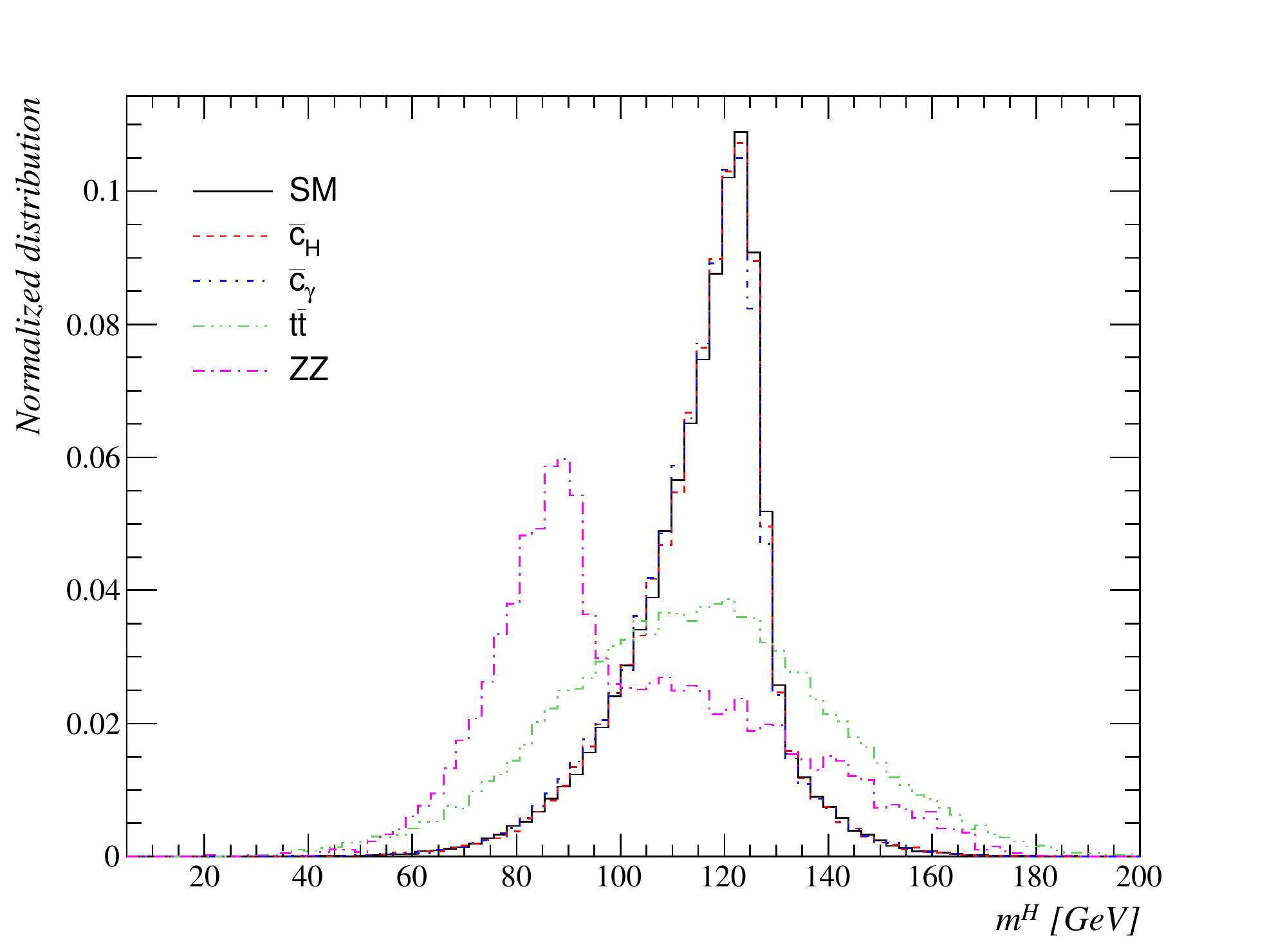}}   
		\resizebox{0.450\textwidth}{!}{\includegraphics{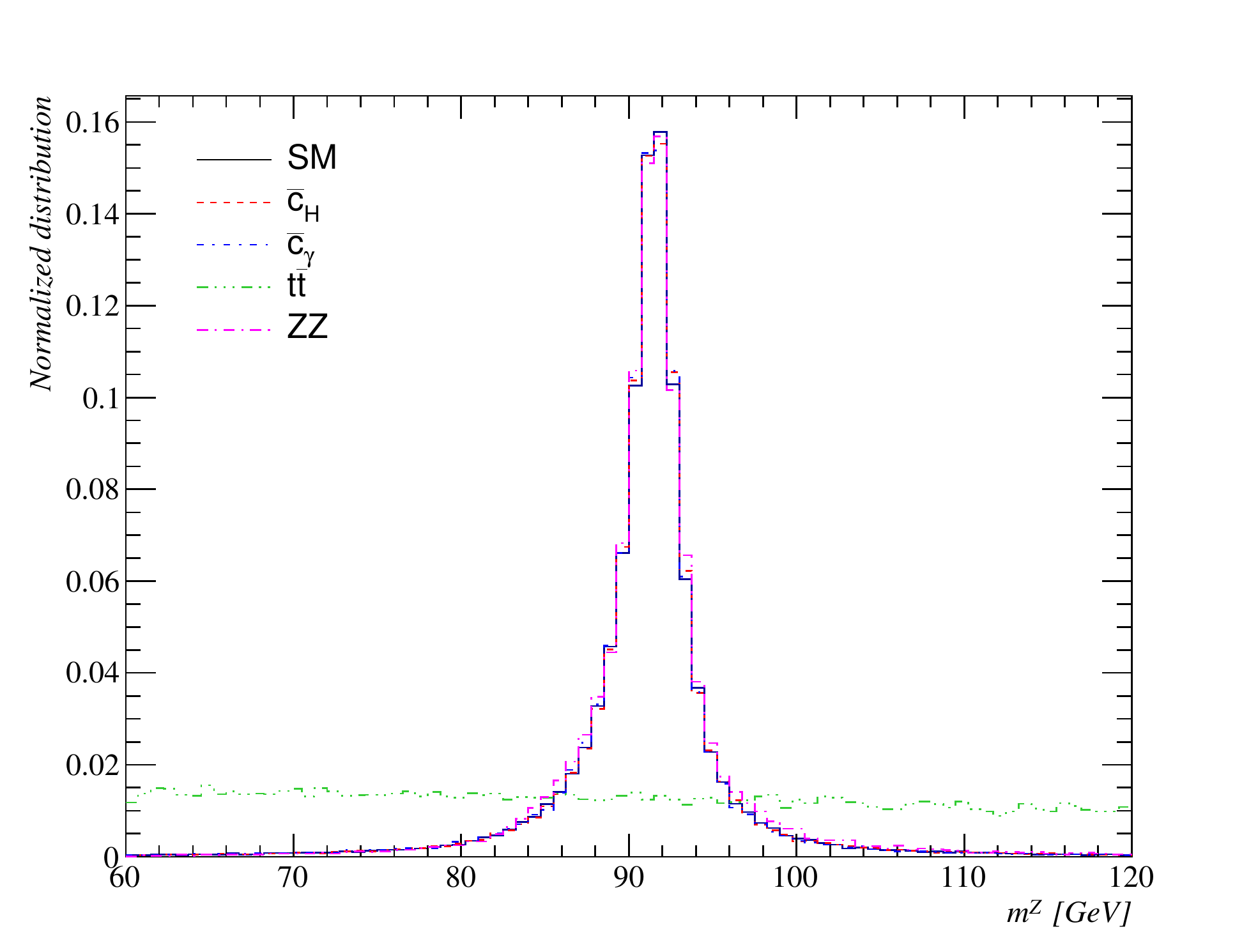}}  
		\caption{ The transverse momentum and pseudo-rapidity, and mass distributions of the
			reconstructed $Z (\ell^{+} \ell^{-})$ and Higgs boson ($b \bar{b}$) for particular values of $\bar{c}_H = 0.1$, $\bar{c}_\gamma = 0.1$ and for SM background processes. The distributions are depicted after the preselection cuts.}\label{fig:PtEta}
	\end{center}
\end{figure*}

To reduce the contributions from all background processes without a Higgs boson or a $Z$ boson
in the final state,  window cuts on the reconstructed Higgs boson and $Z$ boson are applied. It is required 
that $90 < m_{b \bar{b}} < 160$ GeV and $75 < m_{\ell \ell} < 105$ GeV. 
In Figs.~\ref{fig:PtEta}, we  show the transverse momentum, pseudo-rapidity, and the mass distributions of the
reconstructed Higgs ($b\bar{b}$-pair) and $Z$ ($l^{+}l^{-}$-pair) bosons for centre-of-mass energy 
of $\sqrt{s} = 500$ GeV for the SM background processes and for the signal processes with $\bar{c}_H = 0.1$ and $\bar{c}_\gamma = 0.1$.
The distributions are depicted after the preselection cuts.
As it can be seen, the reconstructed Higgs and $Z$ bosons in signal events tend to reside 
at high transverse momentum region while the $t\bar{t}$ background process is in low transverse momentum region.
As a result, the transverse momentum distribution of Higgs or $Z$ bosons are  good variables to suppress the contribution of 
the $t\bar{t}$ background process. In addition to the above selection, an additional cut on the $Z$ boson transverse momentum is applied.
Due to the correlation between the transverse momenta of the $Z$ boson and Higgs boson, only the cut is applied on one of them.

Cross sections of signal and background processes after imposing each set cuts are presented in the Table~\ref{Table:Cut-Table}.
According to Table~\ref{Table:Cut-Table},  the cut on transverse momentum of the reconstructed $Z$ boson ($p_T^{\ell^+ \ell^-}$)
and the window cuts on the reconstructed Higgs and $Z$ boson masses efficiently reject the backgrounds contributions and keep the signal 
events. In particular, these cuts are very useful to reduce the $t \bar{t}$ background process and  $\gamma\gamma,Z\gamma, W^{+} W^{-}Z$ backgrounds. 

\begin{table*}[htbp]
	\begin{center}
		\begin{tabular}{c|c c|c c c c}
			$\sqrt s = 500$ GeV      &  \multicolumn{2}{c|}{Signal }    & \multicolumn{2}{c}{~~~~~~~~~~~~~~~~~~~~~~~ Background }    \\   \hline
			Cuts    & $\bar{c}_H$    & $\bar{c}_\gamma$          & SM ($H+Z$) & $t \bar{t}$ &  $ZZ$   &  $Z\gamma,\gamma\gamma,WWZ$   \\ \hline  \hline
			Cross-sections (in fb)              &  $4.51$  &  $16.76$  & $5.00$        & $24.77$   &  $36.16$     &  $11.47$     \\
(I):	2$\ell$, $|\eta^{\ell}|<2.5$, $p_T^{\ell}>10$    &  $3.41$  &  $12.22$ & $3.79$  &   $15.18$  &  $23.27$  &  $7.37$    \\
(II):	$2{\rm jets}$, $|\eta^{\rm jet}|<2.5$, $p_T^{\rm jet}>20$, $\Delta R_{\ell, \rm jet}\geq 0.5$   &  $2.48$ &  $8.81$   & $2.75$   &  $11.21$   &  $13.95$      &  $4.52$      \\
(III): $2{\rm b-jets}$  &  $1.09$ &  $3.84$   & $1.22$   &  $4.71$   &  $1.16$      &  $0.35$      \\ \hline 			
(IV): $p_T^{\ell^+ \ell^-}>100$    &  $1.06$  &  $3.56$ & $1.18$  &   $0.51$  &  $0.73$  &  $0.094$    \\
(V): $90 <  m_{b \bar{b}} < 160,~75 <  m_{\ell^+ \ell^-} < 105$ &  $0.921$   &  $3.040$   &  $1.022$     &  $0.078$ &   $0.138$       &  $0.003$     \\  \hline  \hline
		\end{tabular}
	\end{center}
	\caption{ Expected cross sections in unit of fb after different combinations of cuts for signal and SM background processes. 
		The signal cross sections are corresponding to particular values of $\bar{c}_H = 0.1$ and  $\bar{c}_\gamma = 0.1$. 
		The center-of-mass energy of the collision is assumed to be 500 GeV. More details of the selection cuts are given in the text. }
	\label{Table:Cut-Table}
\end{table*}

\begin{figure*}[htb]
	\begin{center}
		\vspace{0.50cm}
		\resizebox{0.65\textwidth}{!}{\includegraphics{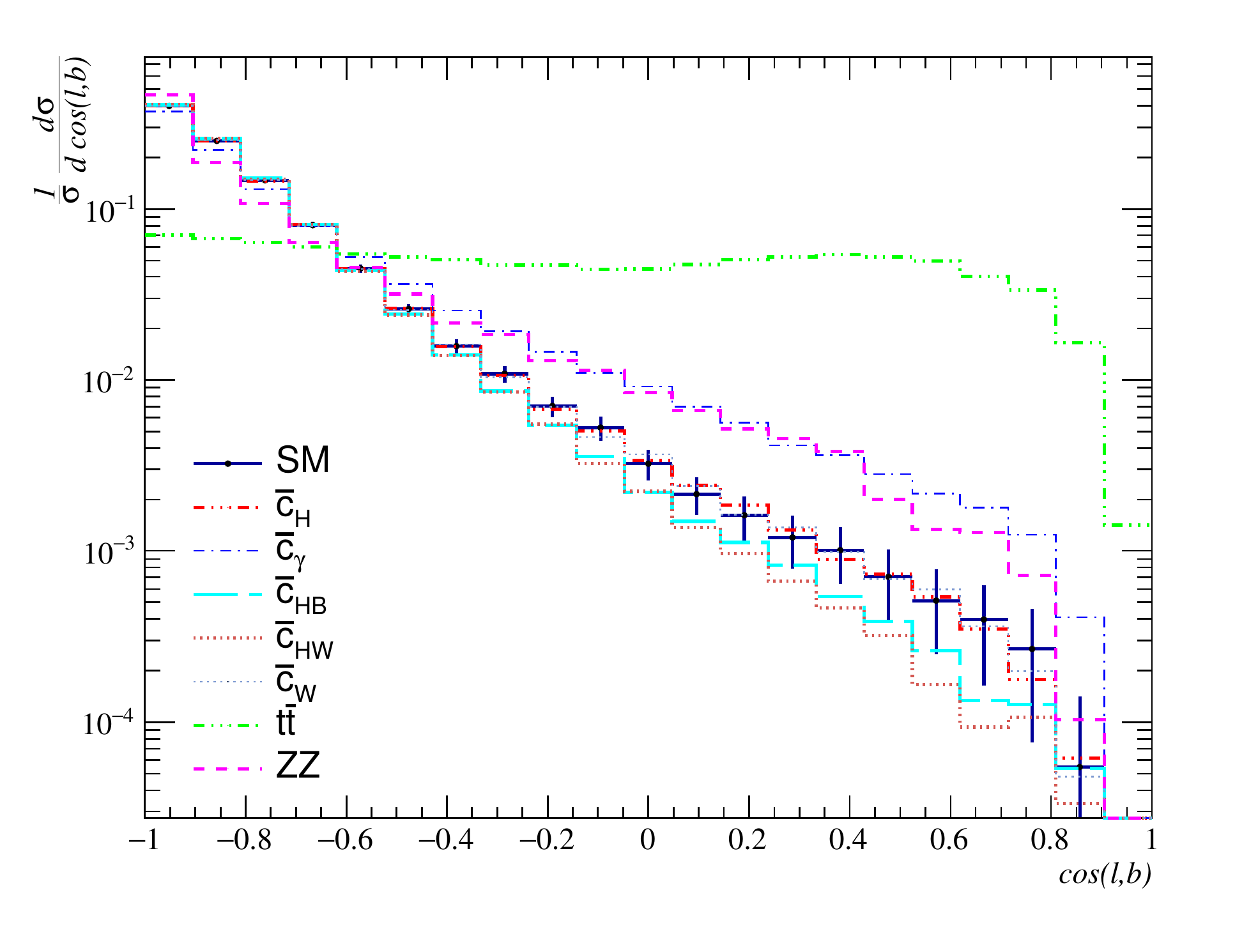}}   
		\caption{ The $\cos (\ell, b-{\rm jets})$ distributions for  SM production of Higgs in association with a $Z$ boson and 
		$H+Z$ production in the present various couplings at the centre-of-mass energy of $\sqrt{s} = 500$ GeV. 
		The distributions are depicted after the preselection cuts.
		The signal distributions are presented for particular values of the coupling set to 0.1. 
		 The distributions of two main background processes $t\bar{t}$ and $ZZ$
		 are depicted for more illustration. The uncertainty on the SM $H+Z$ production is only the statistical uncertainty corresponding to the
		 integrated luminosity of 300 fb$^{-1}$. }\label{fig:Cosleptonbjet}
	\end{center}
\end{figure*}

In order to achieve good sensitivity to the new effective couplings and find the exclusion regions for 
$c_{i}$, a shape analysis on an angular distribution of the final state particles for which 
the shape of signal is different from background processes is performed.
Fig.~\ref{fig:Cosleptonbjet} shows the distribution of the cosine of the angle between the
highest $p_{T}$ b-jet and the highest $p_{T}$ charged lepton, $\cos(\ell,b)$, for signal 
and for the SM background processes after the preselection cuts.
For $H+Z$ signal, the 
charged lepton and b-jet tend to be produced mostly back-to-back at $\cos(\ell,b) \approx -1$.
As it can be seen, some of the dimension six operators could modify the shape 
of the $\cos(\ell,b)$ distribution with respect to the SM production of Higgs associated with a $Z$ boson.
 For example, switching on $\bar{c}_{\gamma}$ leads to increase the 
number of events in the region of $\cos(\ell,b) > 0$ while non-zero value of $\bar{c}_{HW}$
leads to decrease the number of events in the region of $\cos(\ell,b) > 0$ with respect to 
the SM Higgs production in association with a $Z$ boson. 
The $t\bar{t}$  background process has  almost a flat distribution
while $ZZ$ process has a shape almost similar to the $\bar{c}_{\gamma}$ signal process.

At this point, it should be mentioned that other distributions such as the
Higgs boson transverse momentum which differentiates between signal and 
background processes could be used to derive limits on the
new couplings. In particular,  performing a simultaneous likelihood fit
on both distributions ($\cos(\ell,b)$ and $p_{T}^{H}$) would lead to better results. In the present work, 
only the $\cos(\ell,b)$ distributions of signal and background processes
are used to obtain the upper limits on the new effective couplings.

%
\section{  Statistical method }\label{sec:statistical}
%

In order to obtain exclusion regions in the $(\bar{c}_{i}; \bar{c}_{j})$ plane, where $\bar{c}_{i,j}$ are coefficients
of the dimension six operators defined in Eq.~\ref{eq:L}, a binned $\chi^{2}$ analysis 
is performed on the $d\sigma/dcos(\ell, b)$ distribution.
At a time, we switch on  two effective couplings ($\bar{c}_{i}; \bar{c}_{j}$)
as well as the SM $H+Z$ process and all background processes with the same final
state. Therefore, the $\chi^{2}$ is a function of two effective couplings ($\bar{c}_{i}; \, \bar{c}_{j}$)
and has the following form:

\begin{eqnarray}
	\chi^{2}(\bar{c}_{i}; \, \bar{c}_{j}) = \sum_{i}^{n_{bins}}\Big[\frac{N^{\rm th}_{i}(\bar{c}_{i}; \, \bar{c}_{j}) - N^{\rm exp}_{i}}{\Delta N^{\rm exp}_{i}}   \Big]^{2},
\end{eqnarray}
where $N^{\rm th}_{i}(\bar{c}_{i}; \, \bar{c}_{j}) = \sigma_{i}(\bar{c}_{i}; \, \bar{c}_{j}) \times \epsilon_{i} \times B(H\rightarrow b \bar{b})\times \mathcal{L}$ and $N^{\rm exp}_{i}$
are the number of signal and SM expected events in $i$th bin of the 
$\cos(\ell, b)$ distribution. $\sigma_{i}(\bar{c}_{i}; \, \bar{c}_{j})$ is the cross section of the signal process in $i$th bin of the 
$\cos(\ell, b)$ distribution  and $\mathcal{L}$ is the integrated luminosity.  The selection efficiency in each bin is denoted by 
$\epsilon_{i}$ and $B(H\rightarrow b \bar{b})$ is the branching fraction of Higgs boson decay into $b\bar{b}$ pair in the SM framework. For $m_{H} = 125$ GeV, the value of the branching fraction of Higgs boson decay into $b\bar{b}$ is 
$0.584$ with the relative theoretical uncertainty of $+0.032$ and $-0.033$ \cite{Olive:2016xmw}. 
The combined statistical and systematic uncertainties in each bin is denoted
by $\Delta N^{\rm exp}_{i}$. It is defined as: $\Delta N^{\rm exp}_{i} = \sqrt{N^{\rm SM+bkg}_{i}(1+\Delta_{\rm sys}^{2}\times N^{\rm SM+bkg}_{i})}$
where $\Delta_{\rm sys}$ reflects the effect of an overall systematic uncertainty. In this work, the results are presented 
with and without considering any systematic effects. The predicted constraints at 95\% CL considering only one Wilson coefficient in the 
above fit are obtained as well.

%
\section{  Analysis results }\label{sec:results}
%

\begin{figure*}[htb]
	\begin{center}
		\vspace{0.5cm}
		\resizebox{0.24\textwidth}{!}{\includegraphics{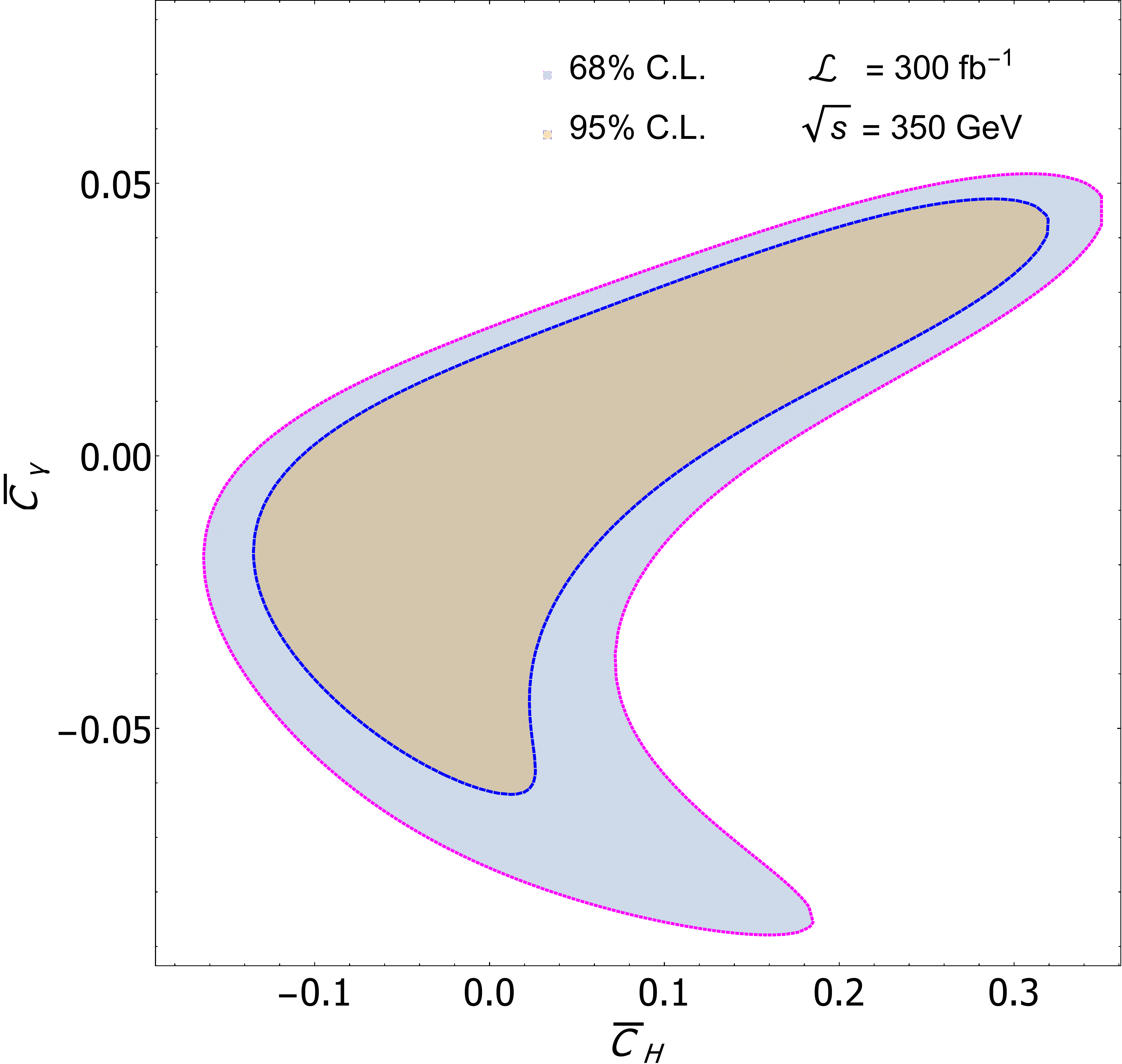}}   
		\resizebox{0.24\textwidth}{!}{\includegraphics{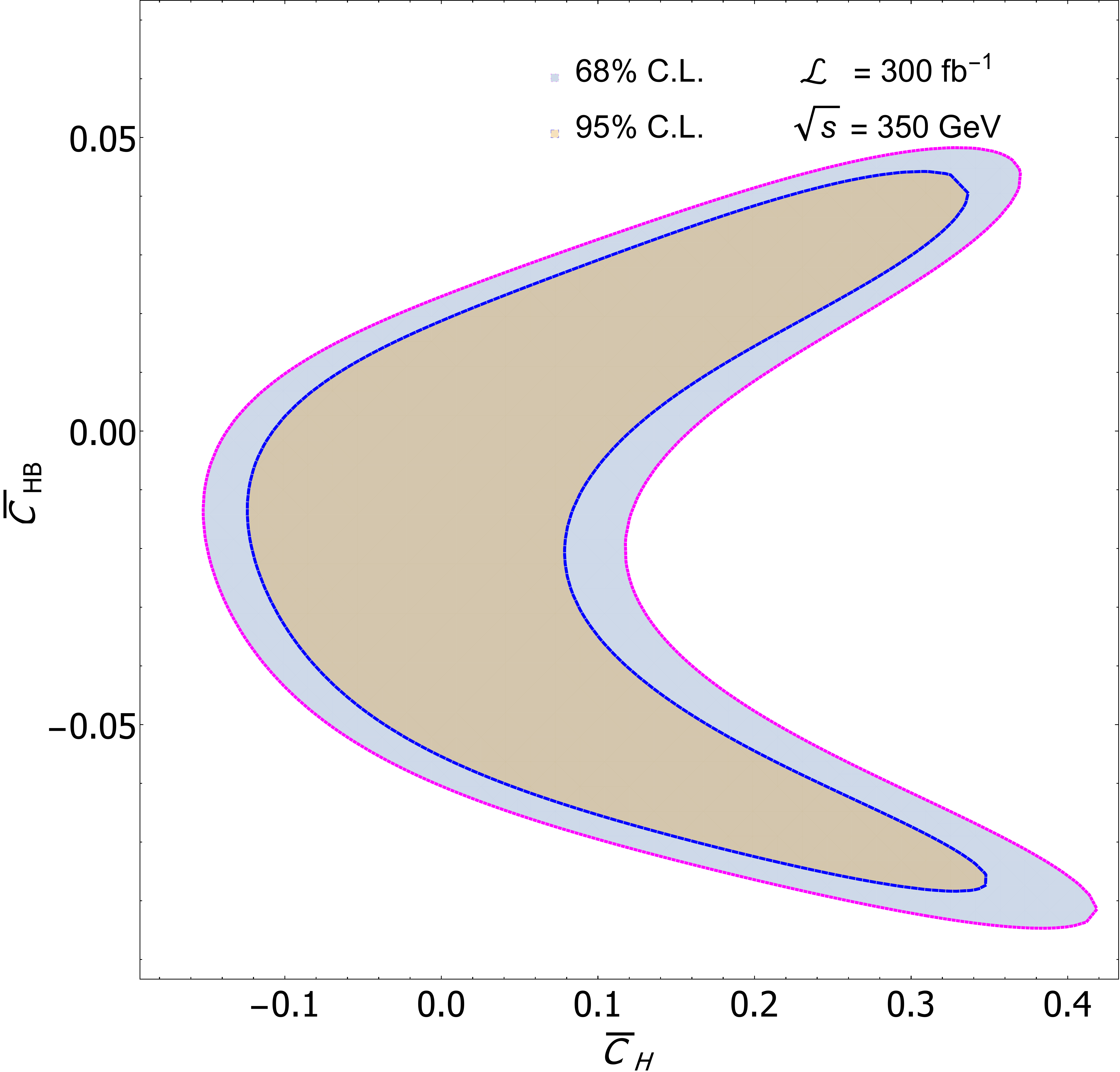}}   
		\resizebox{0.24\textwidth}{!}{\includegraphics{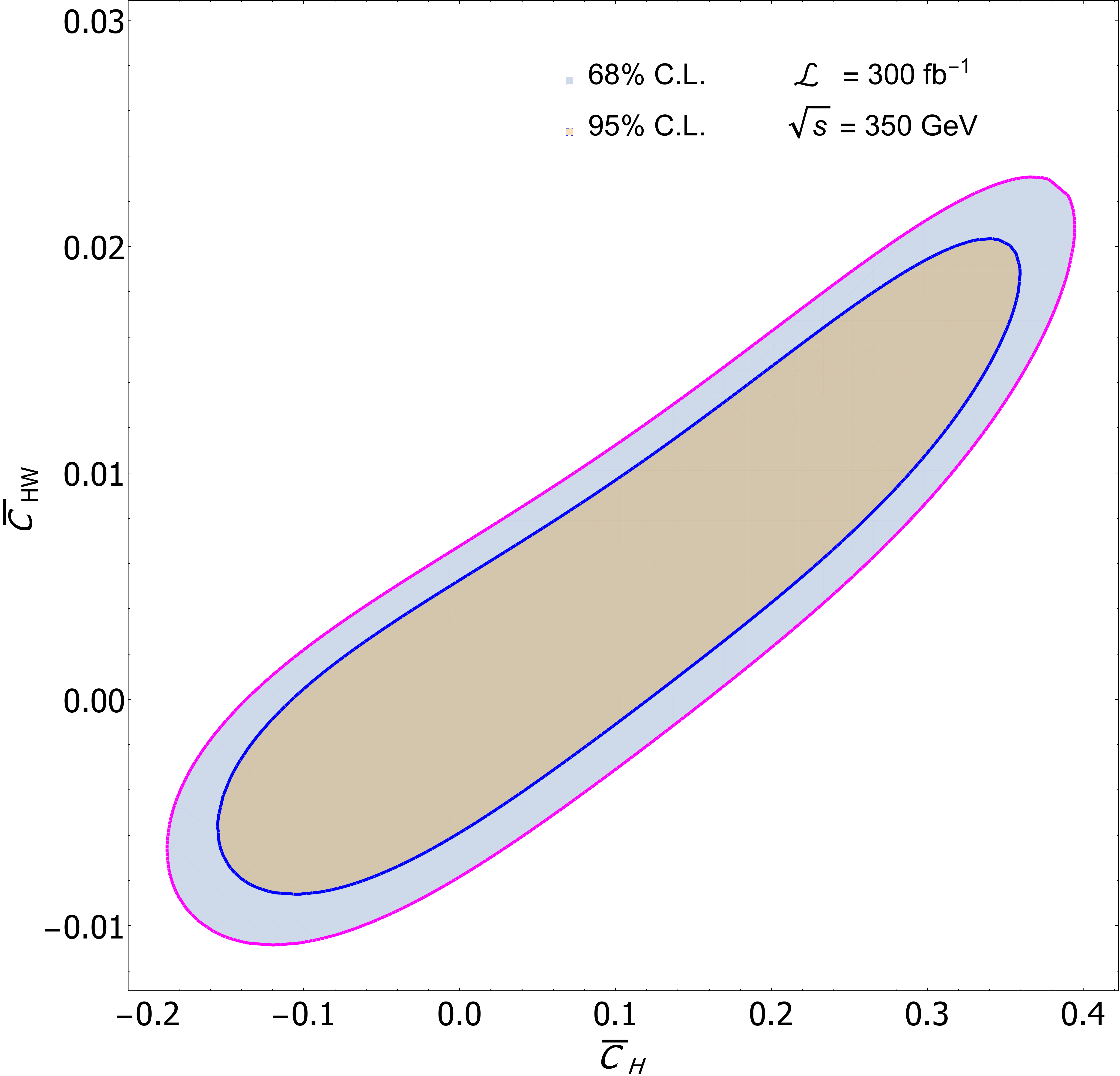}}   
		\resizebox{0.24\textwidth}{!}{\includegraphics{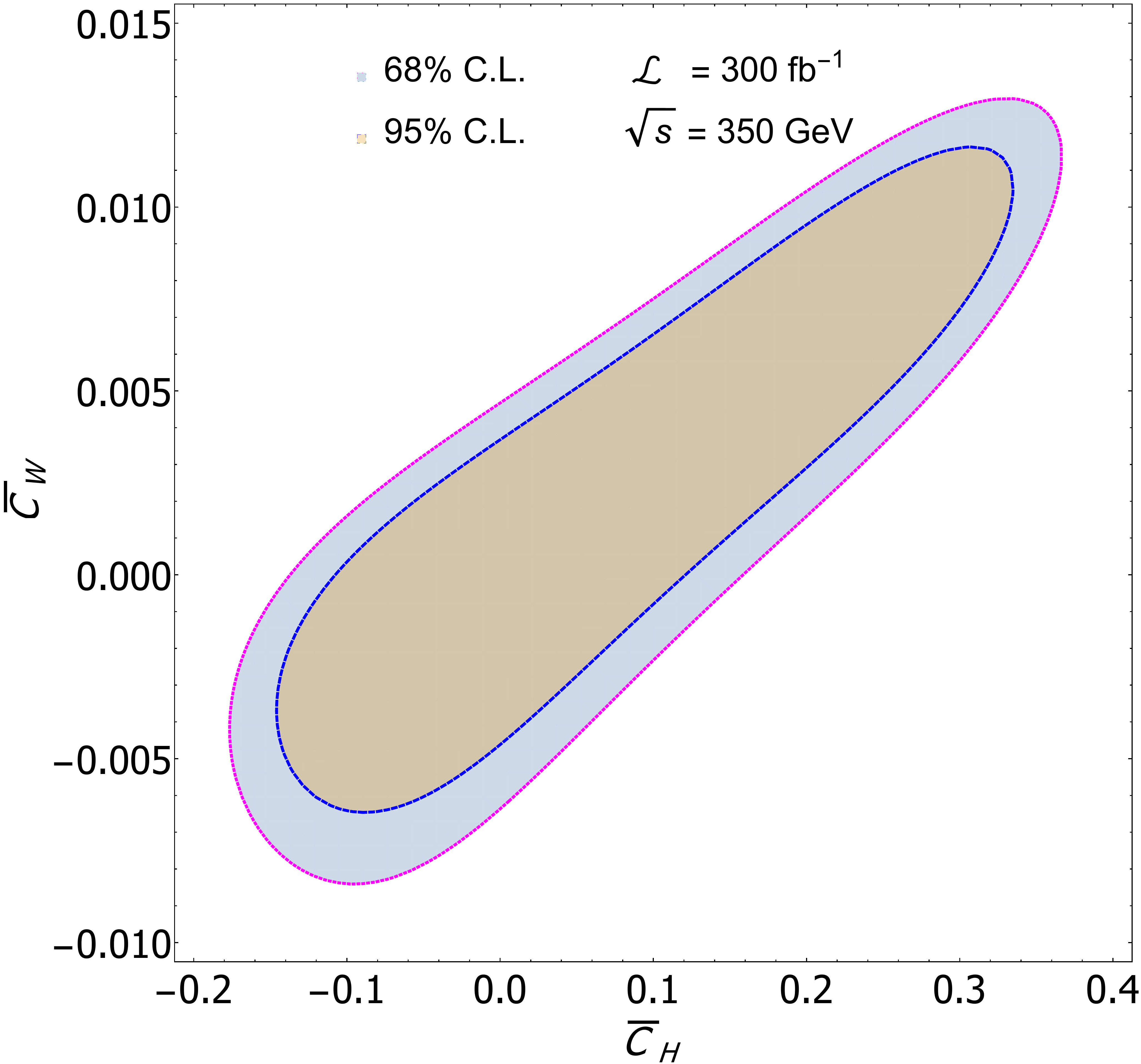}}   
		\vspace{0.5cm}		
		\resizebox{0.24\textwidth}{!}{\includegraphics{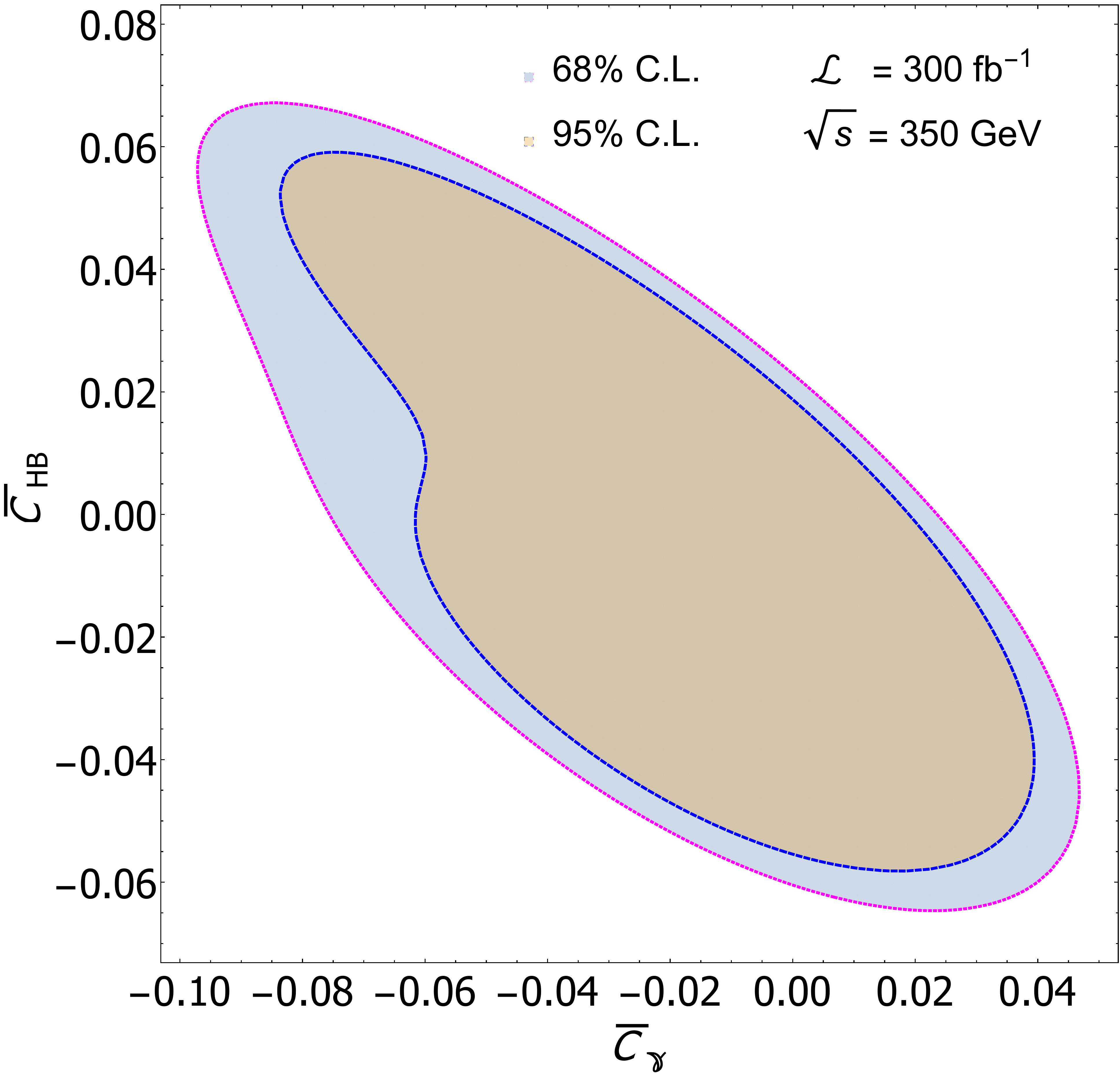}}   
		\resizebox{0.24\textwidth}{!}{\includegraphics{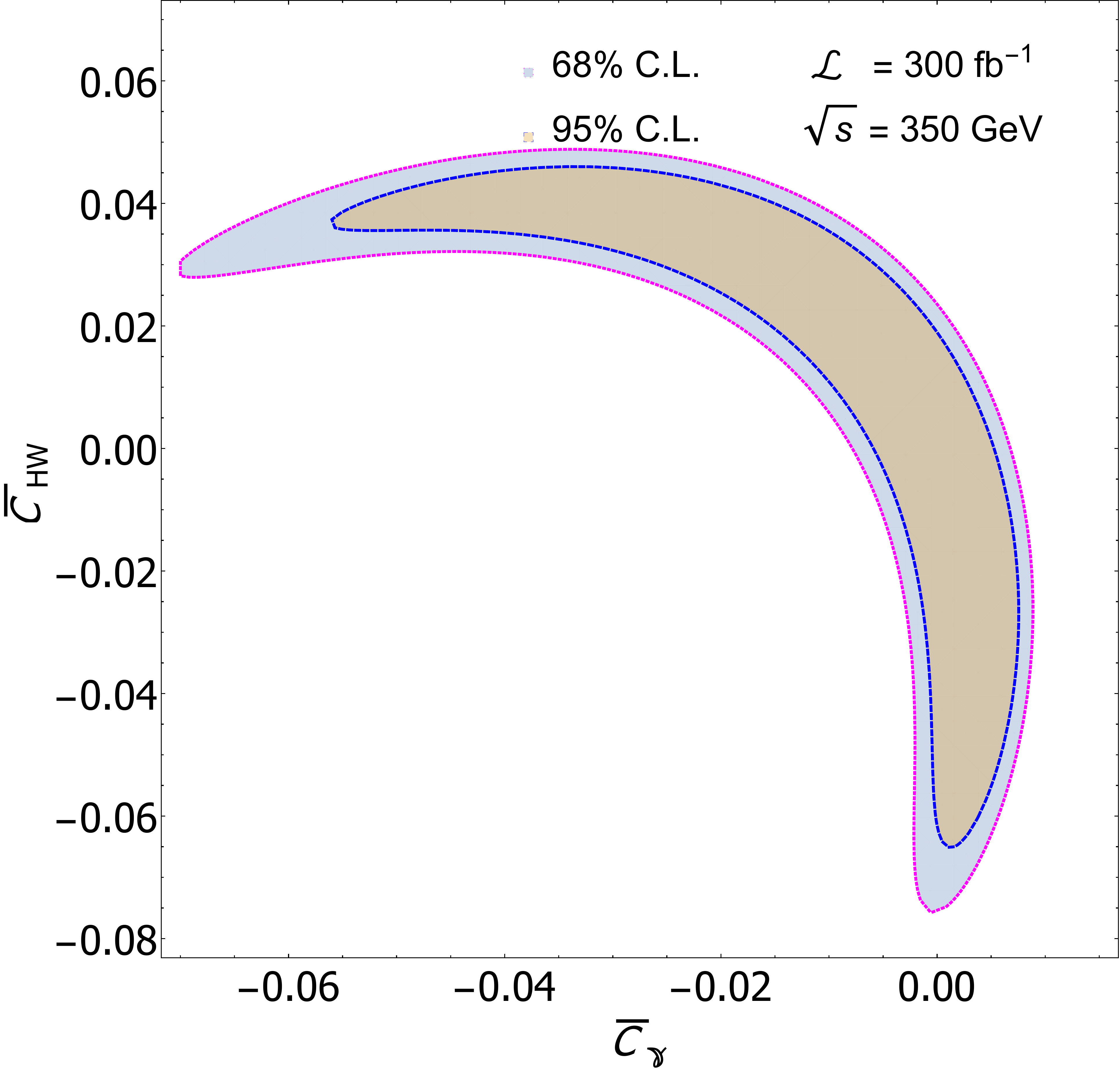}}   
		\resizebox{0.24\textwidth}{!}{\includegraphics{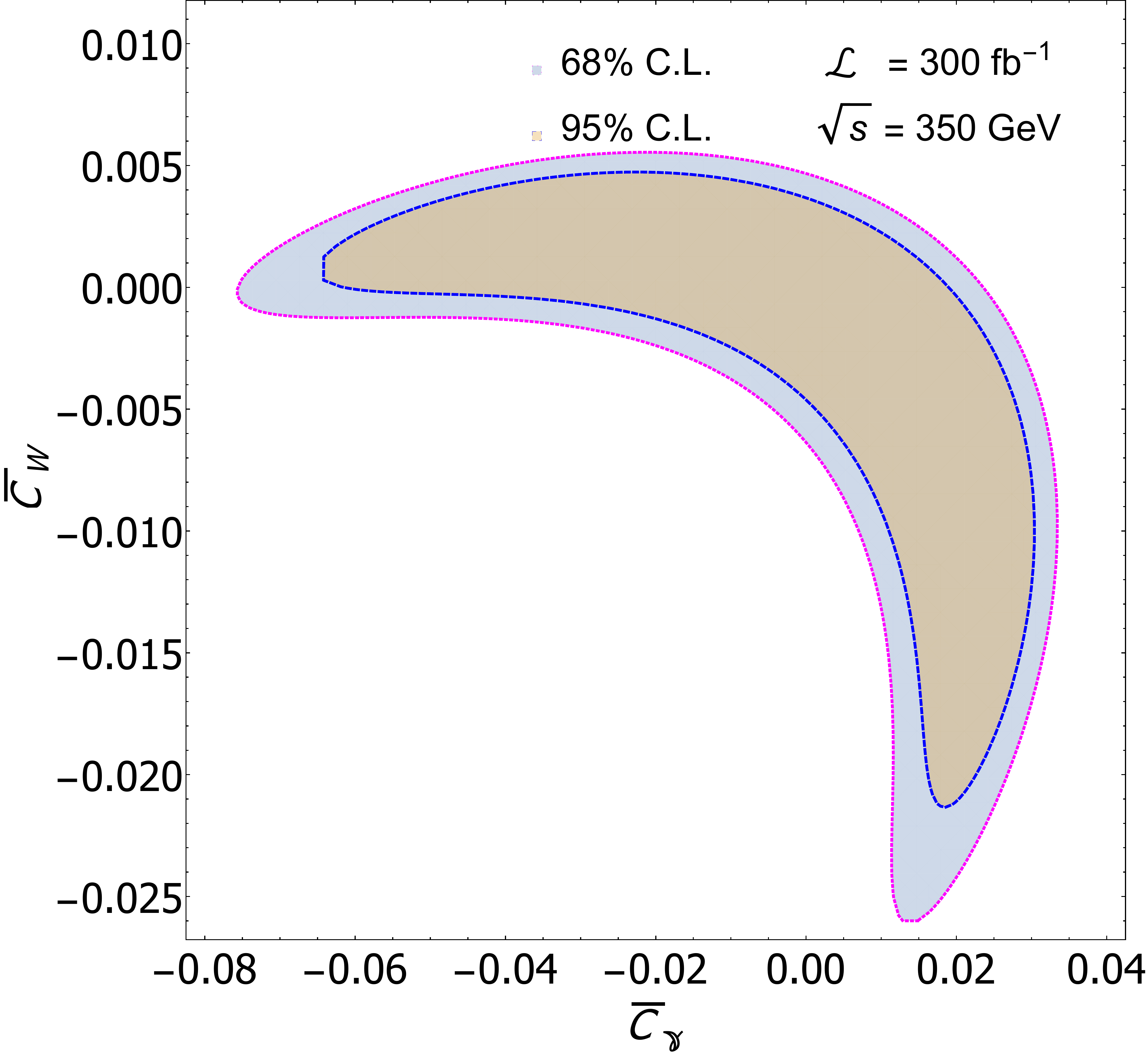}}   
		\vspace{0.5cm}		
		\resizebox{0.24\textwidth}{!}{\includegraphics{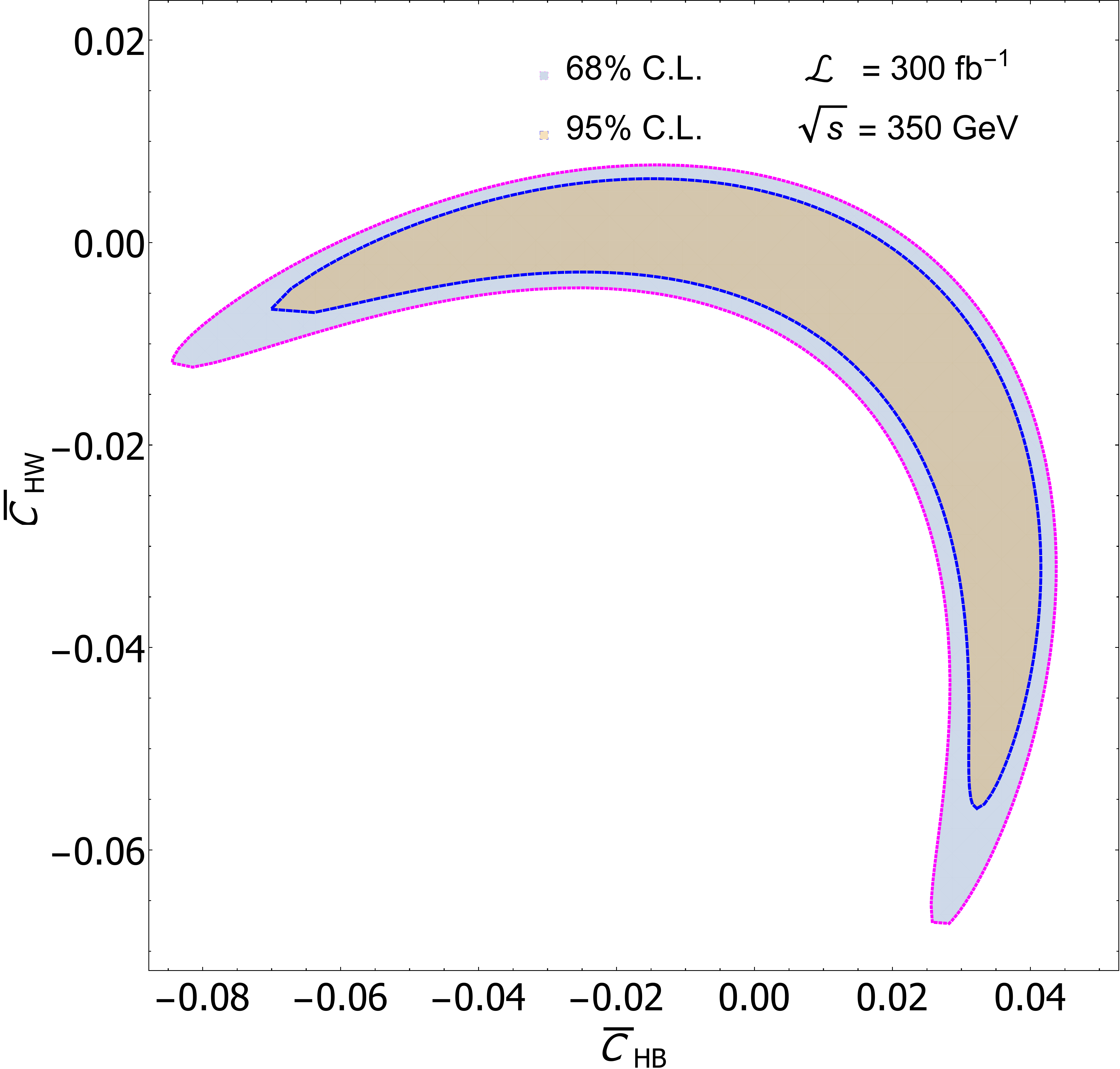}}   
		\resizebox{0.24\textwidth}{!}{\includegraphics{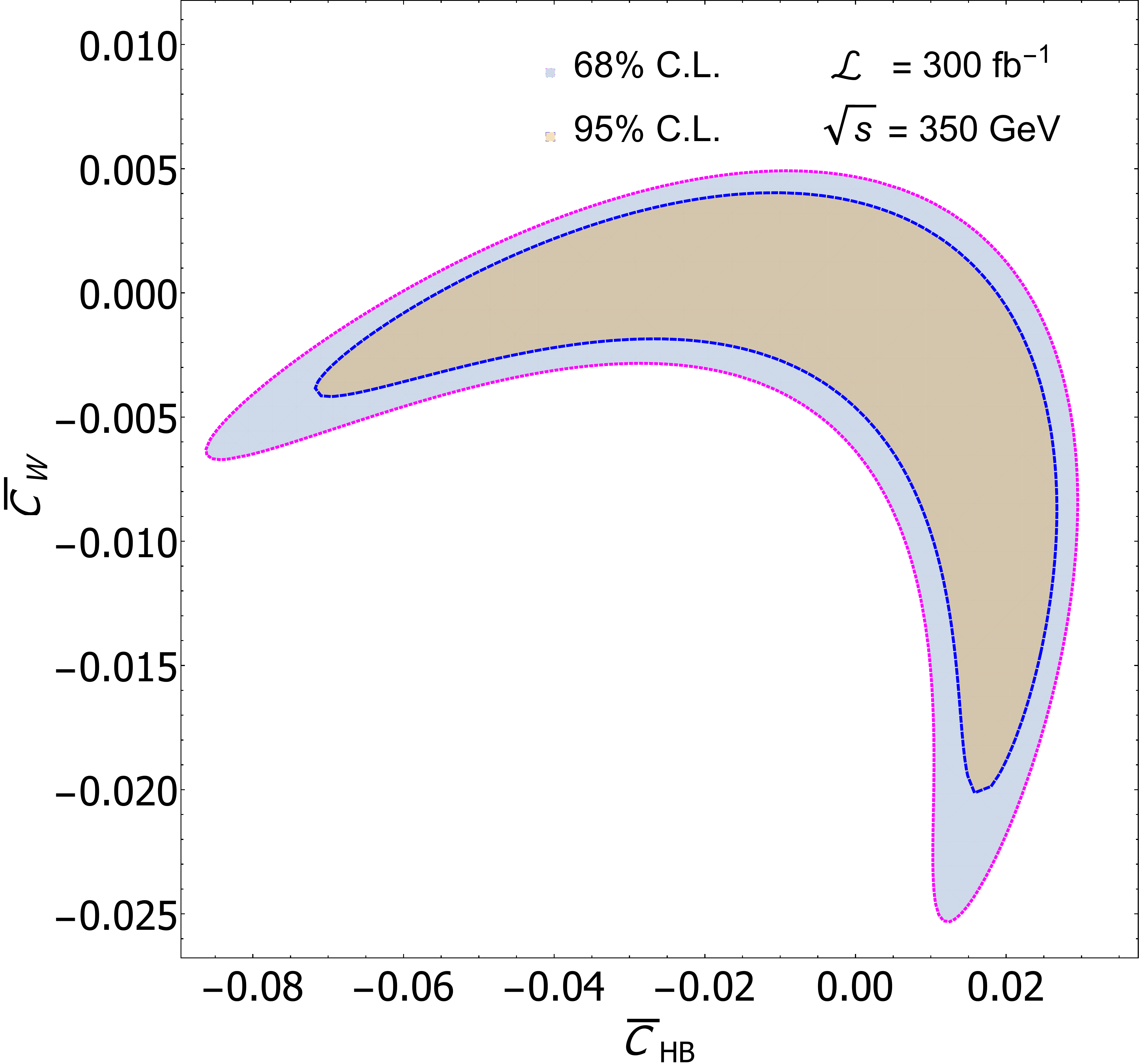}}   
		\vspace{0.5cm}		
		\resizebox{0.24\textwidth}{!}{\includegraphics{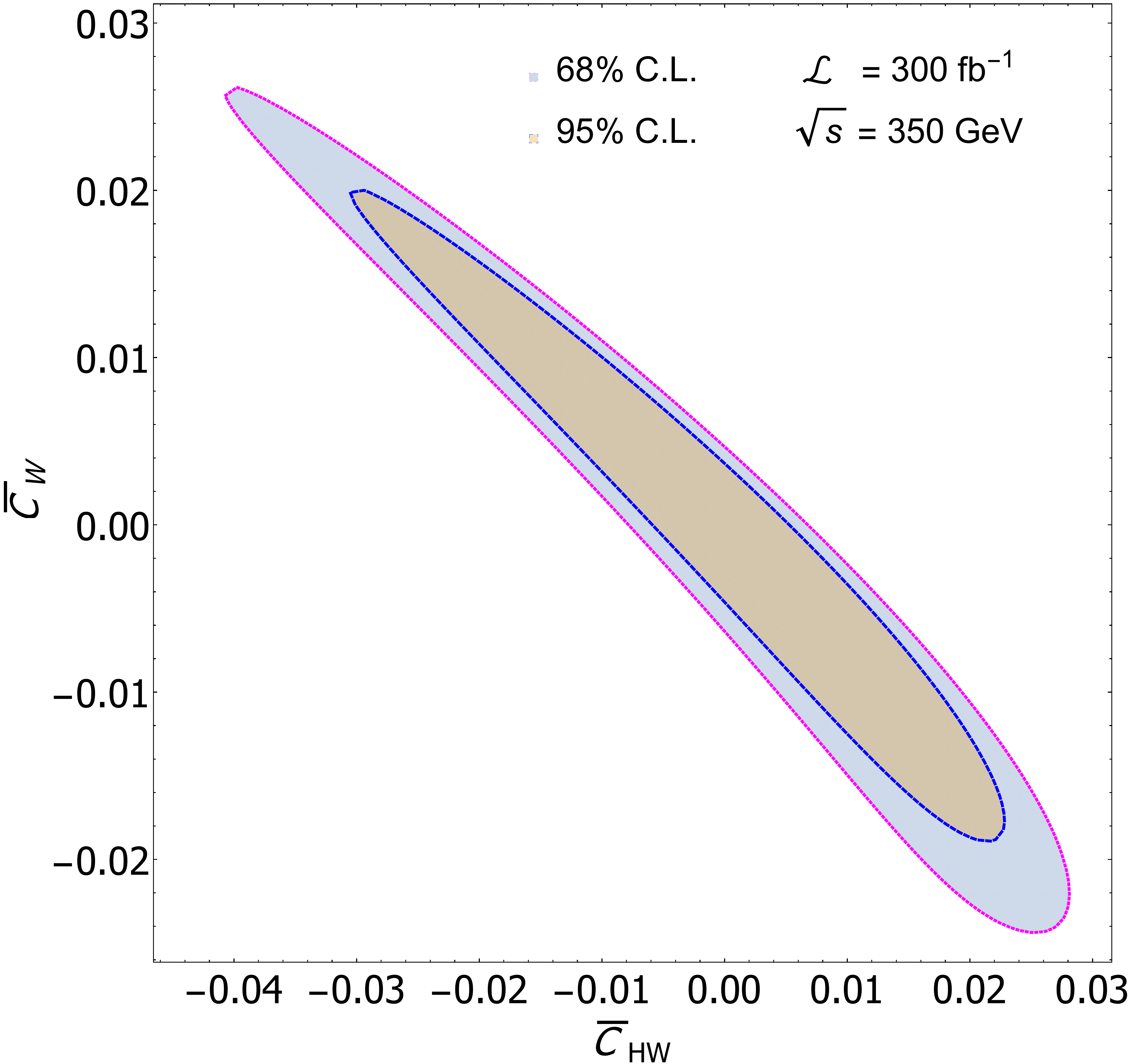}}   
		\caption{ Contours of 68\% and 95\% confidence level obtained from a fit using the
			$\cos (\ell, b-jets)$ distributions for $\sqrt{s} = 350$ GeV with a luminosity of 300 fb$^{-1}$.  }\label{fig:X1-eeHZ-350GeV}
	\end{center}
\end{figure*}

\begin{figure*}[htb]
	\begin{center}
		\vspace{0.5cm}
		\resizebox{0.24\textwidth}{!}{\includegraphics{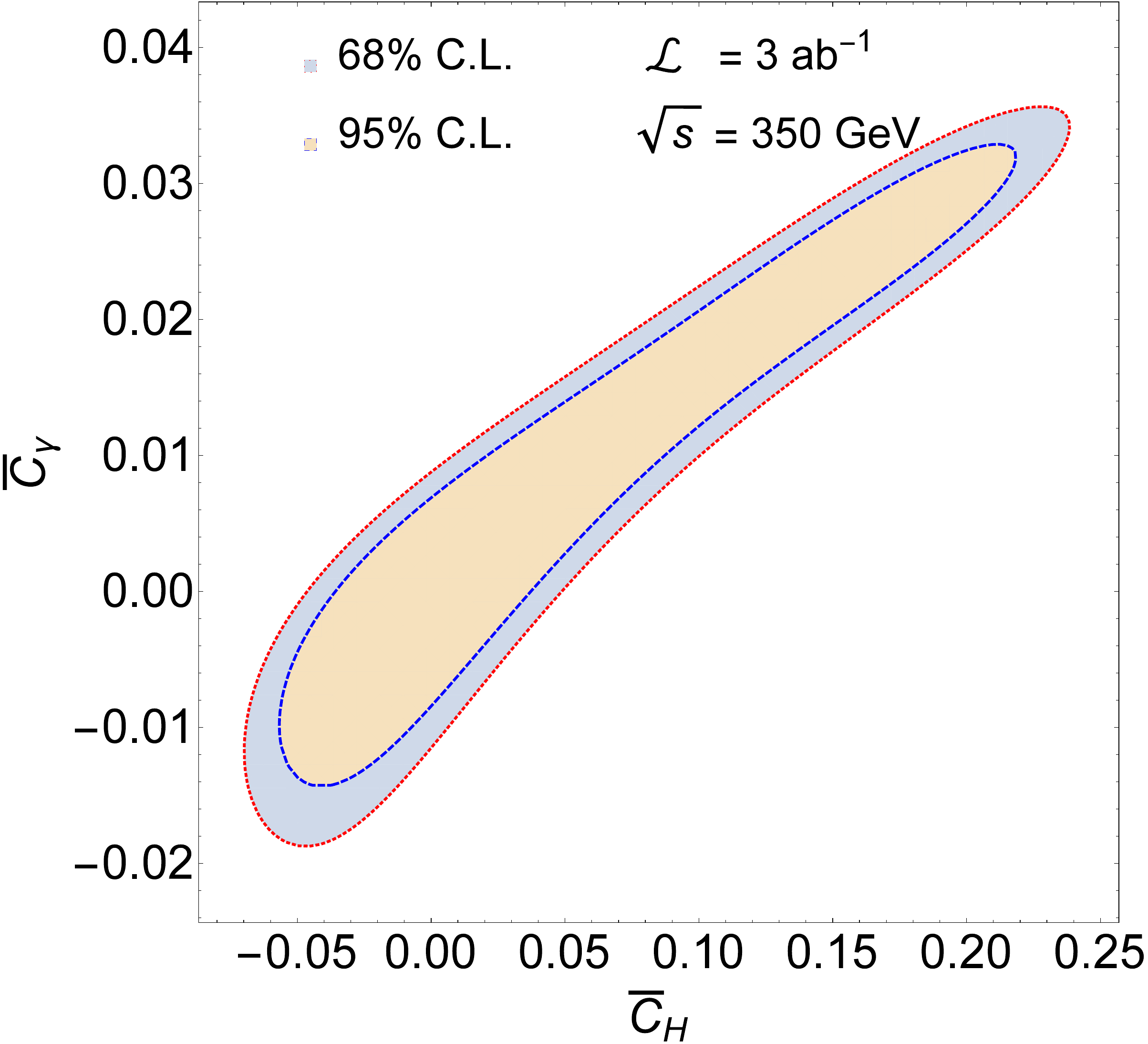}}   
		\resizebox{0.24\textwidth}{!}{\includegraphics{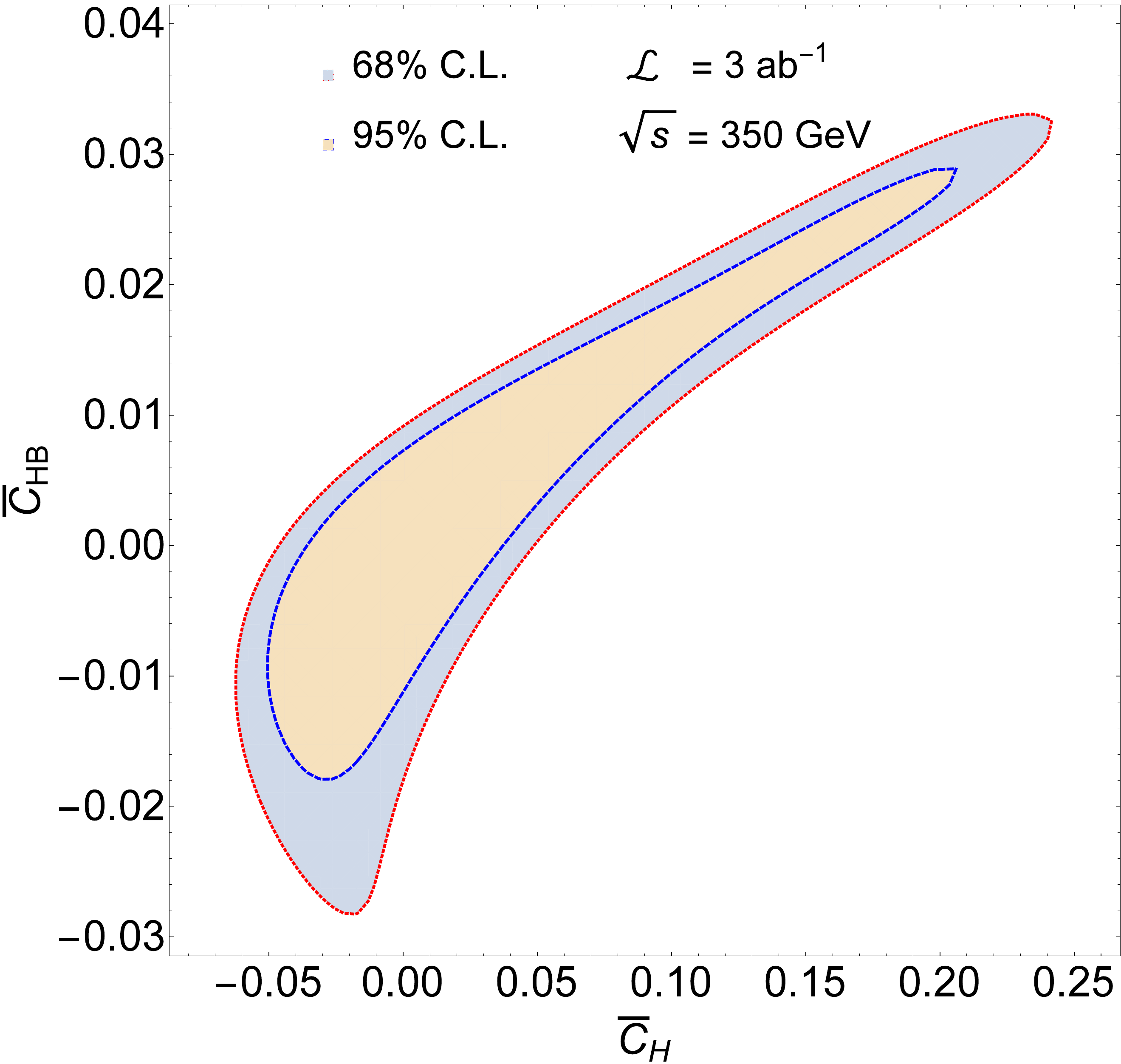}}   
		\resizebox{0.24\textwidth}{!}{\includegraphics{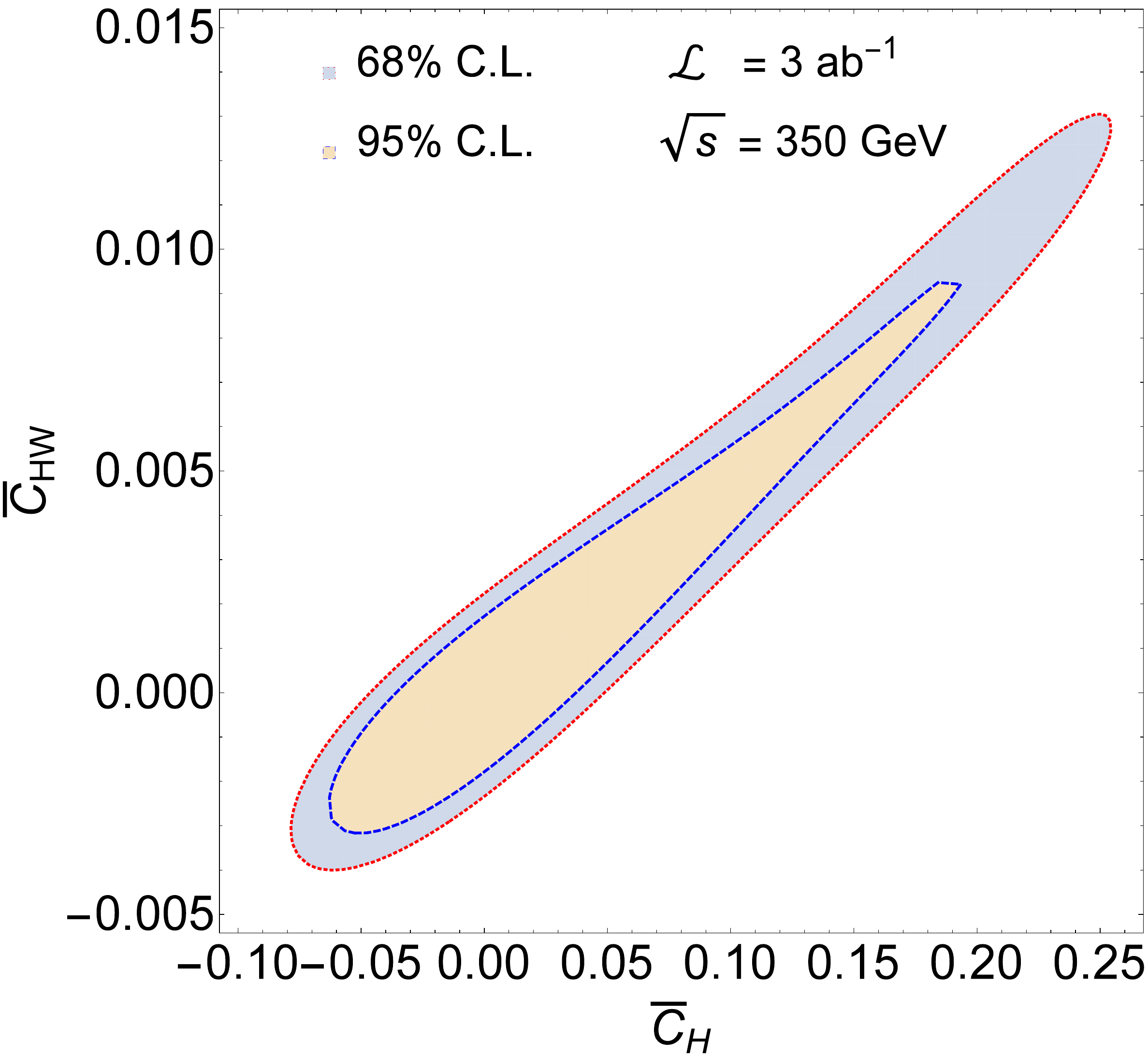}}   
		\resizebox{0.24\textwidth}{!}{\includegraphics{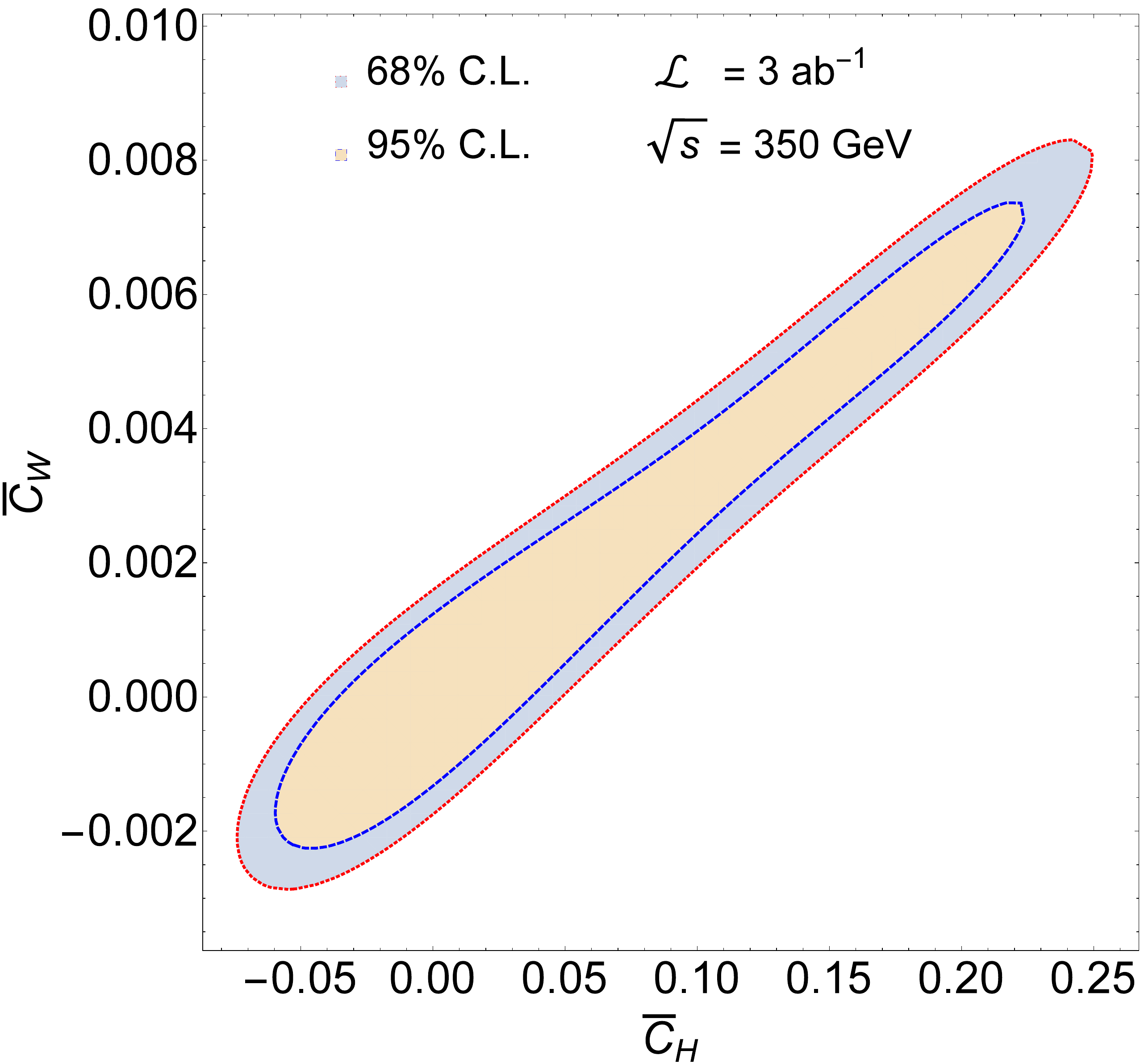}}   
		\vspace{0.5cm}		
		\resizebox{0.24\textwidth}{!}{\includegraphics{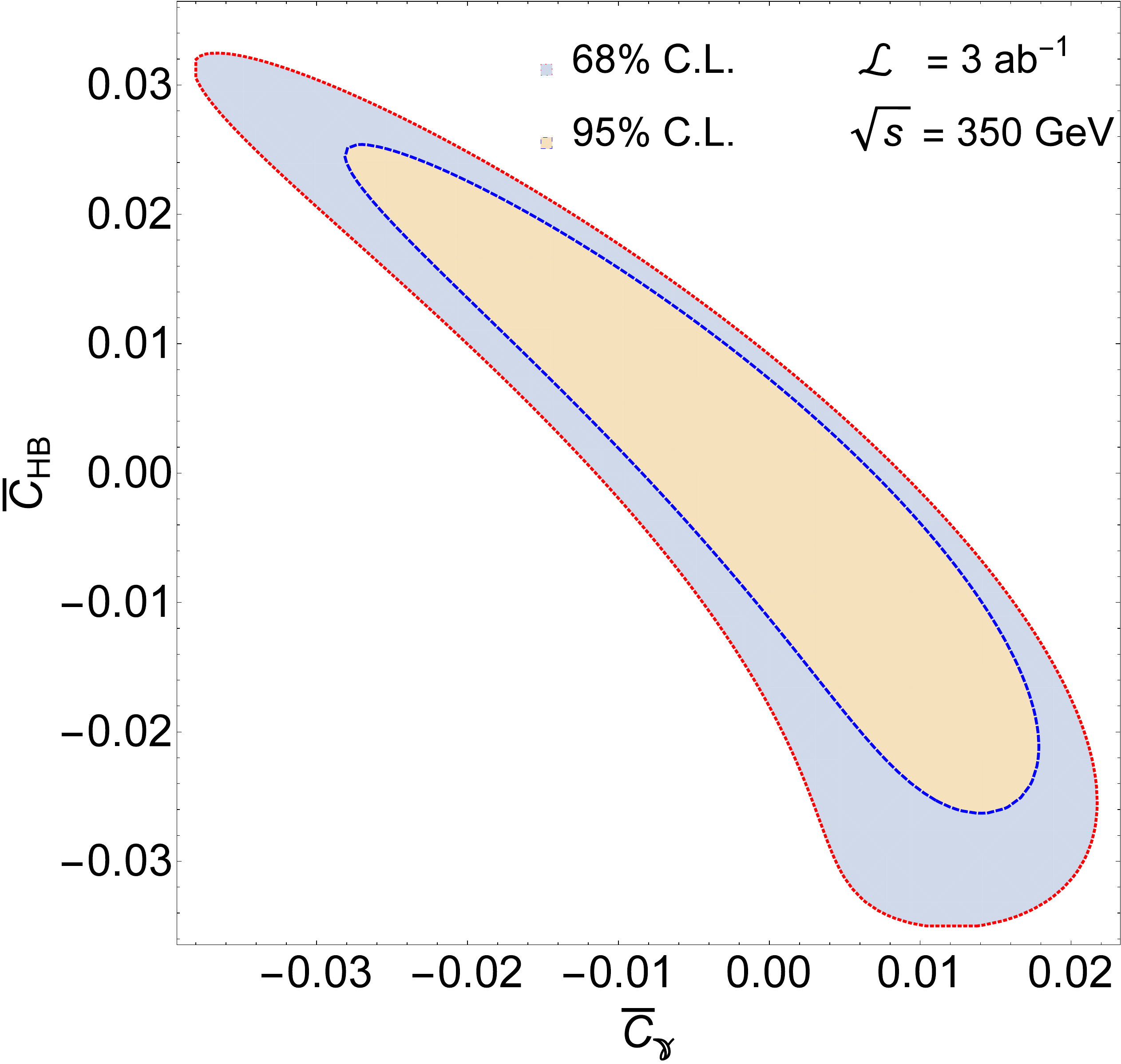}}   
		\resizebox{0.24\textwidth}{!}{\includegraphics{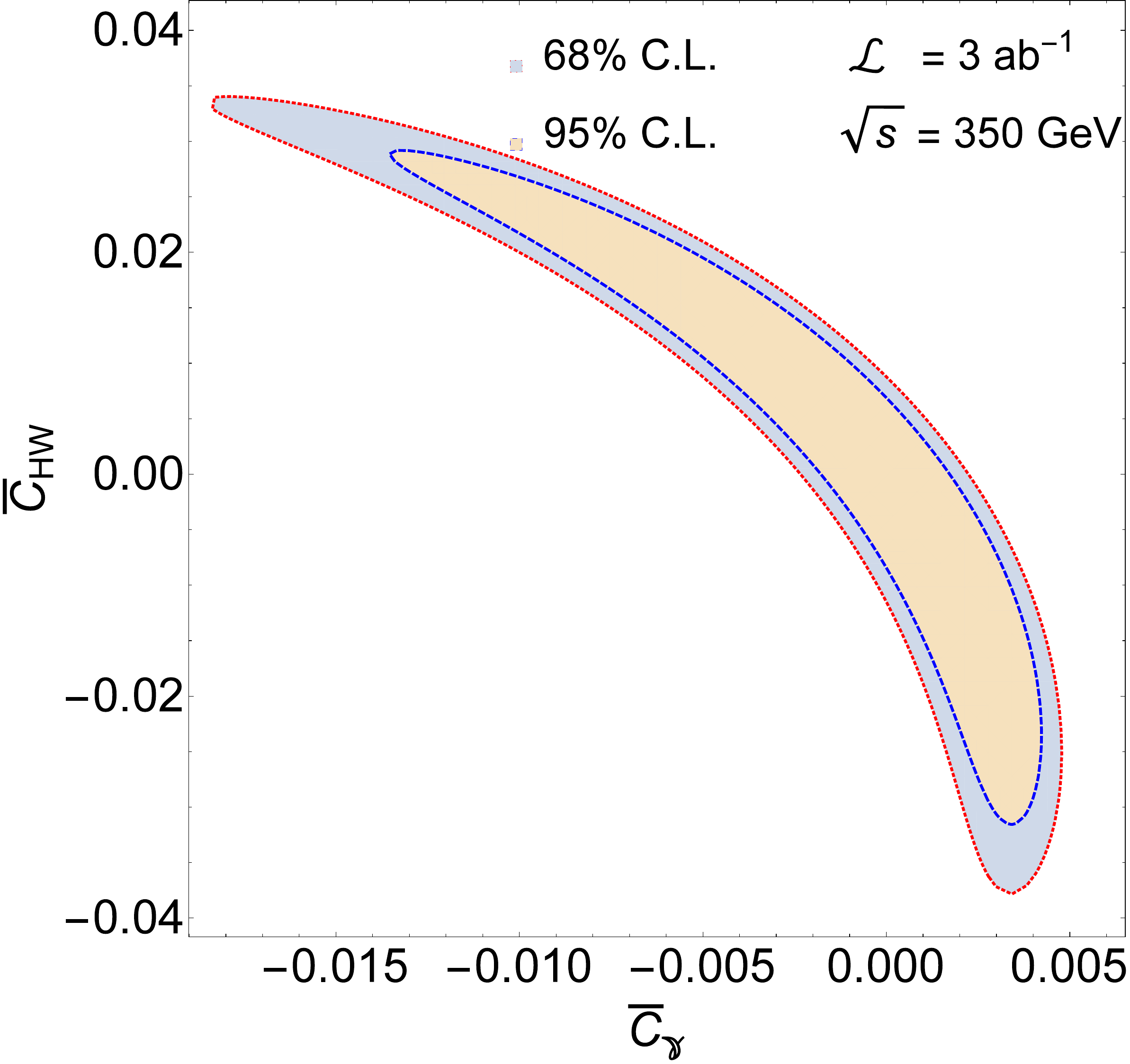}}   
		\resizebox{0.24\textwidth}{!}{\includegraphics{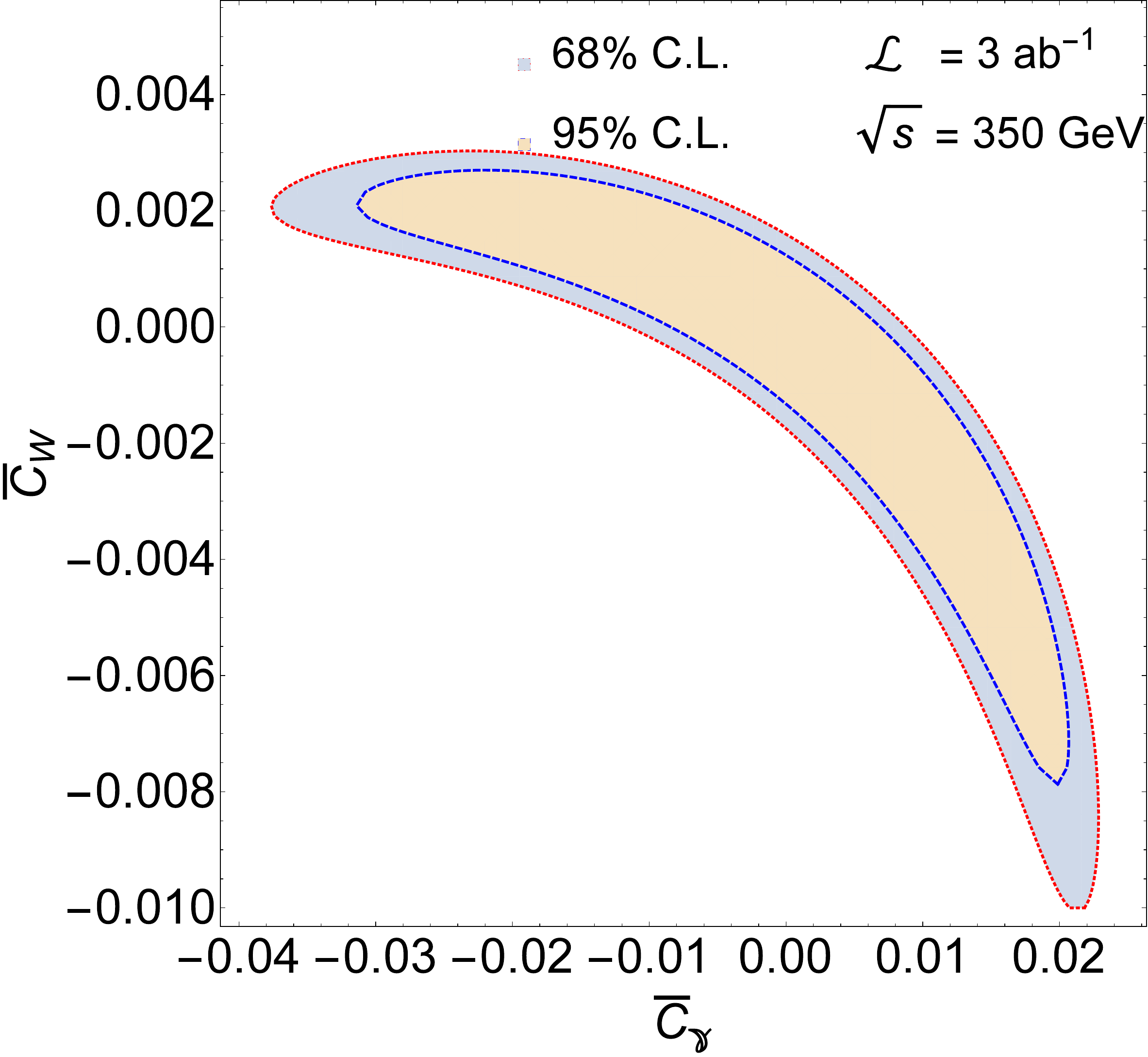}}   
		\vspace{0.5cm}		
		\resizebox{0.24\textwidth}{!}{\includegraphics{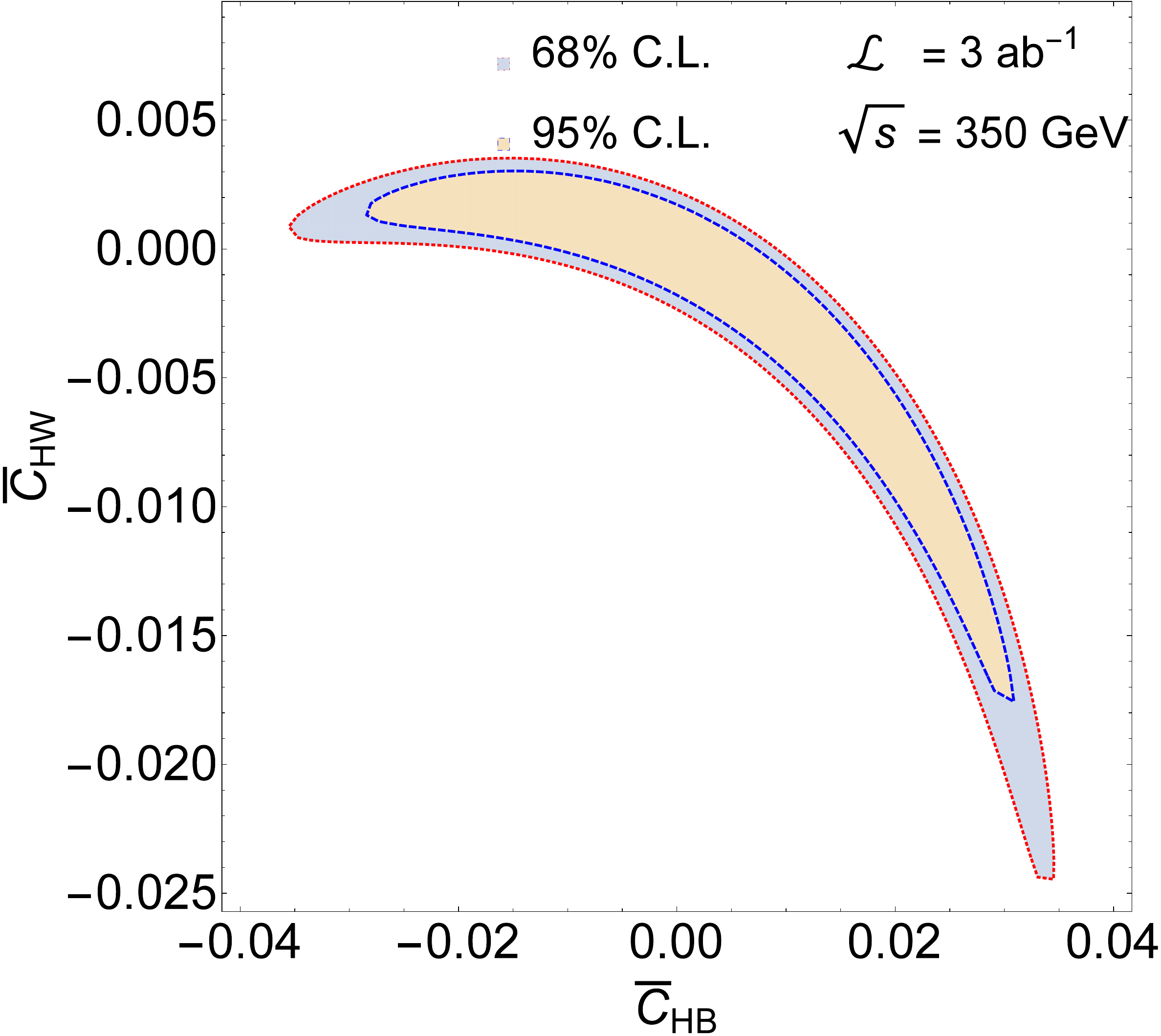}}   
		\resizebox{0.24\textwidth}{!}{\includegraphics{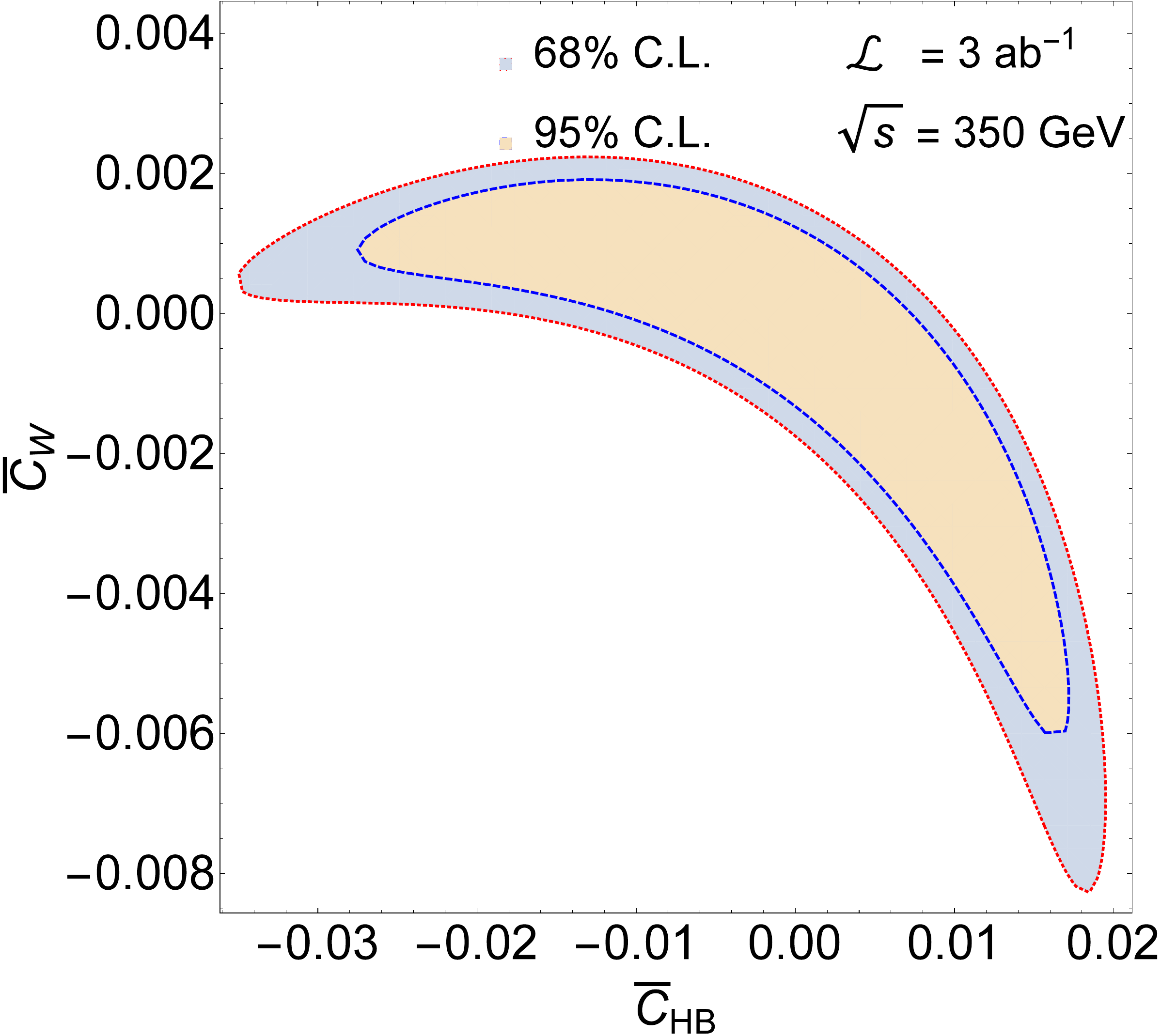}}   
		\vspace{0.5cm}		
		\resizebox{0.24\textwidth}{!}{\includegraphics{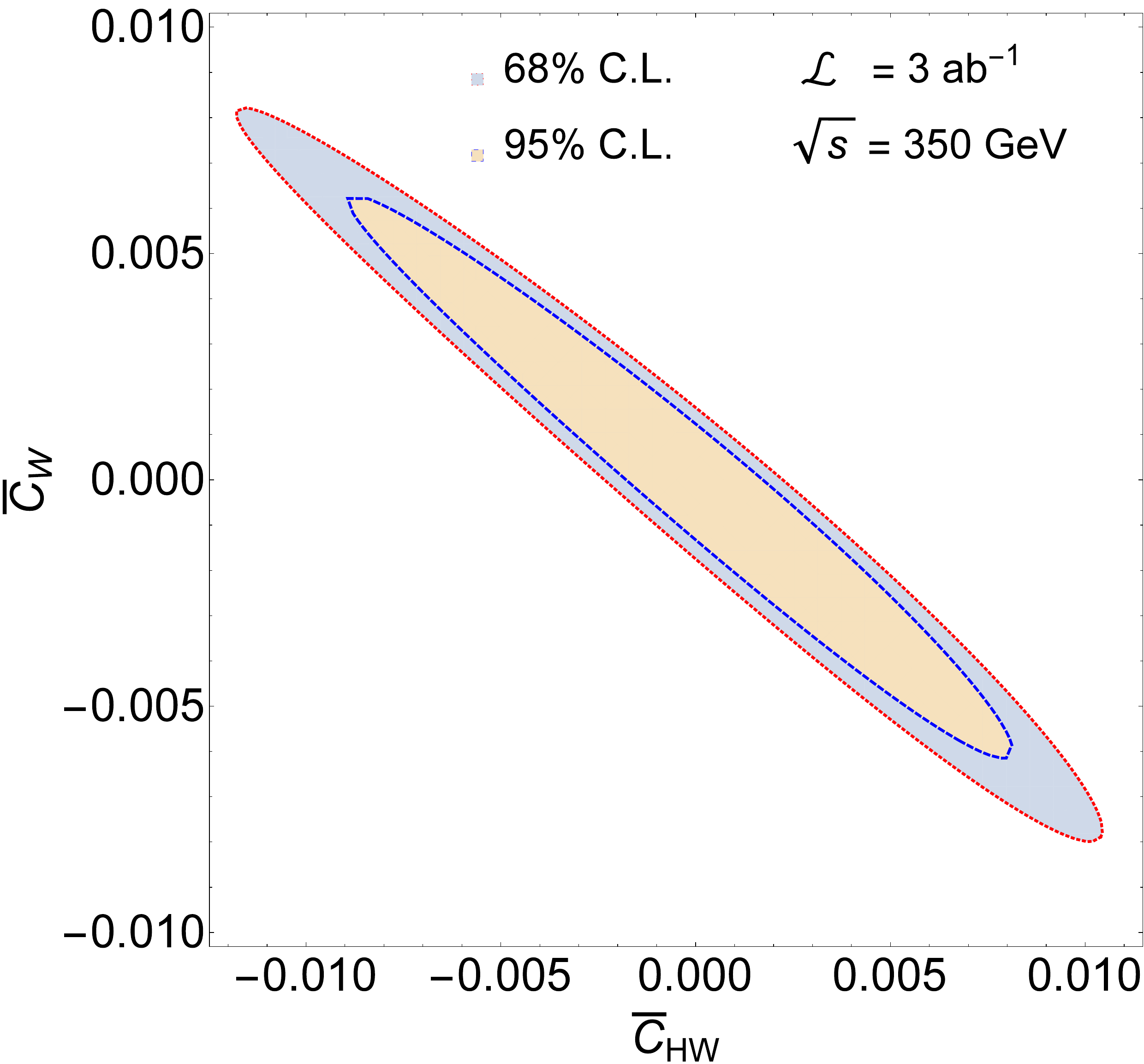}}   
		\caption{ Contours of 68\% and 95\% confidence level obtained from a fit using the
			$\cos (\ell, b-jets)$ distributions for $\sqrt{s} = 350$ GeV with an integrated luminosity of 3 ab$^{-1}$.  }\label{fig:X2-eeHZ-350GeV}
	\end{center}
\end{figure*}

\begin{figure*}[htb]
	\begin{center}
		\vspace{0.5cm}
		\resizebox{0.24\textwidth}{!}{\includegraphics{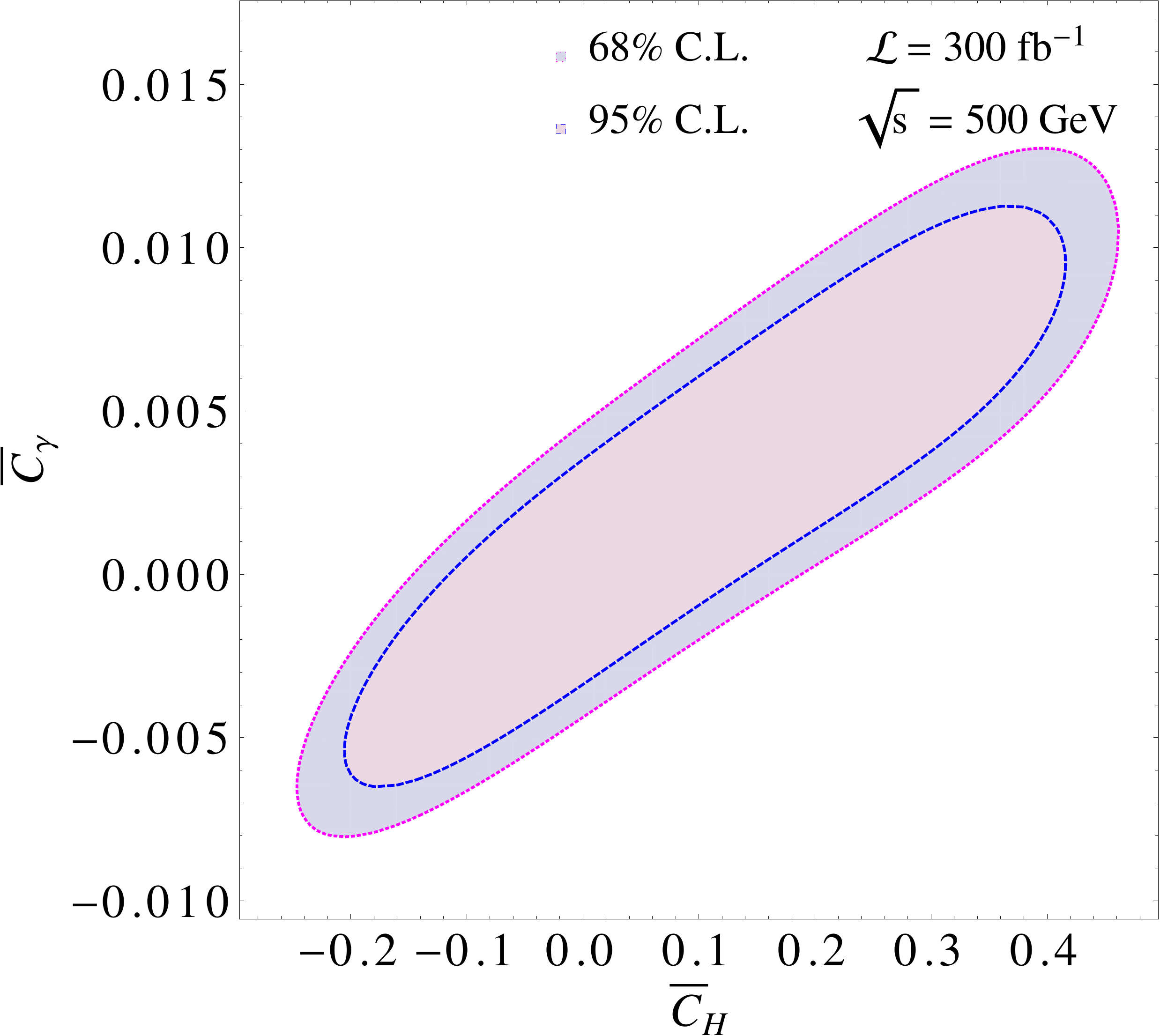}}   
		\resizebox{0.24\textwidth}{!}{\includegraphics{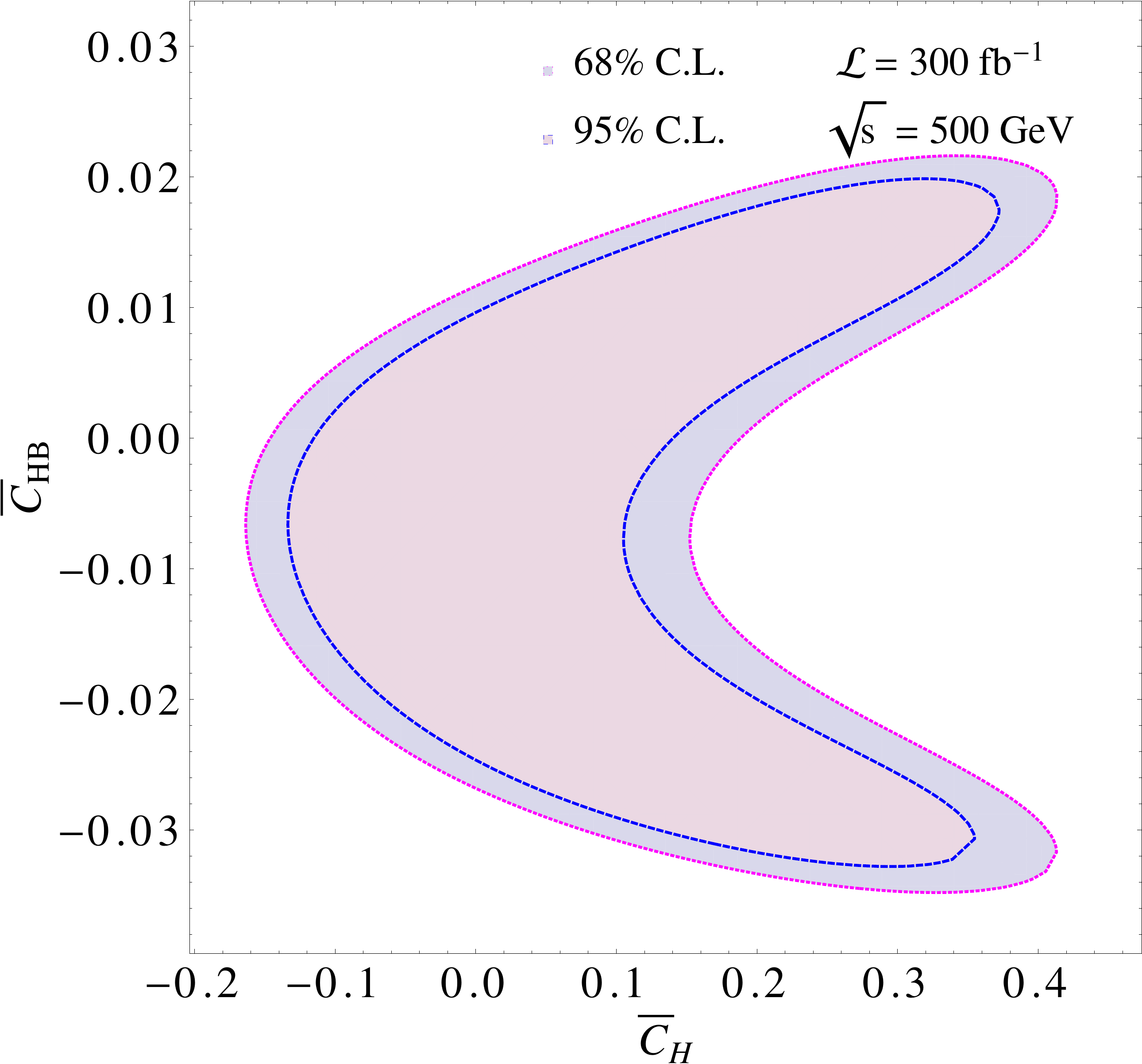}}   
		\resizebox{0.24\textwidth}{!}{\includegraphics{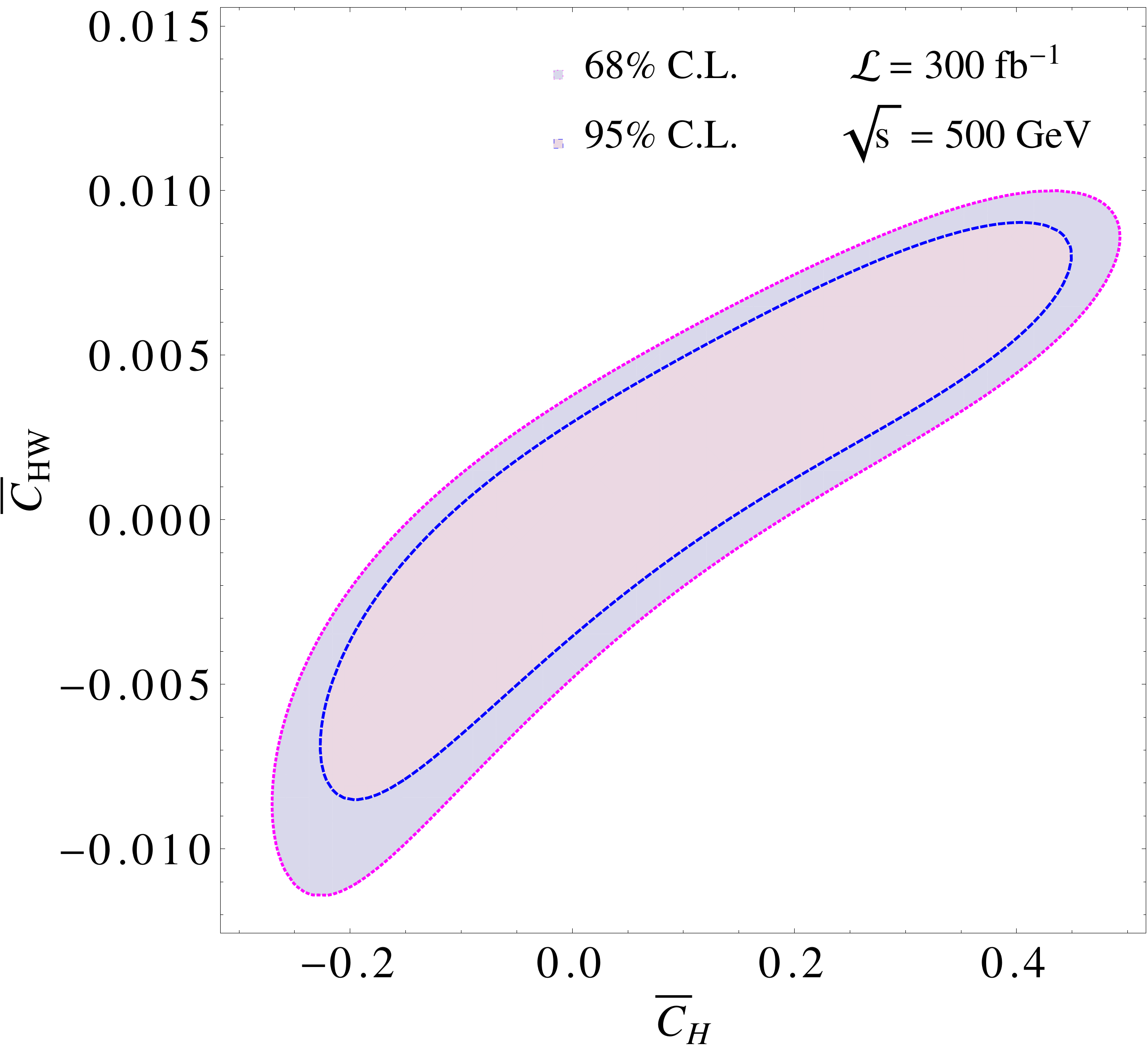}}   
		\resizebox{0.24\textwidth}{!}{\includegraphics{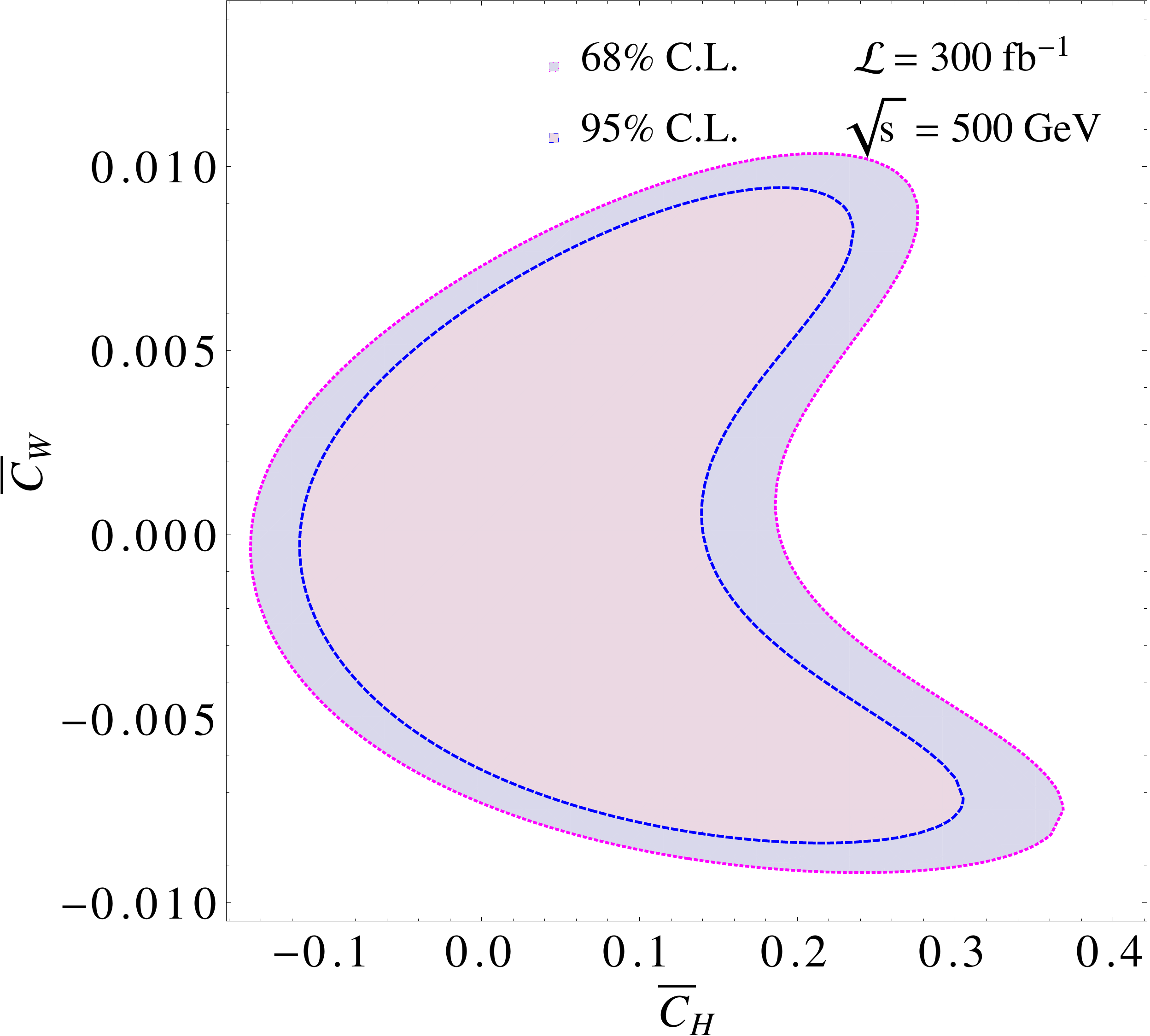}}   
		\vspace{0.5cm}		
		\resizebox{0.24\textwidth}{!}{\includegraphics{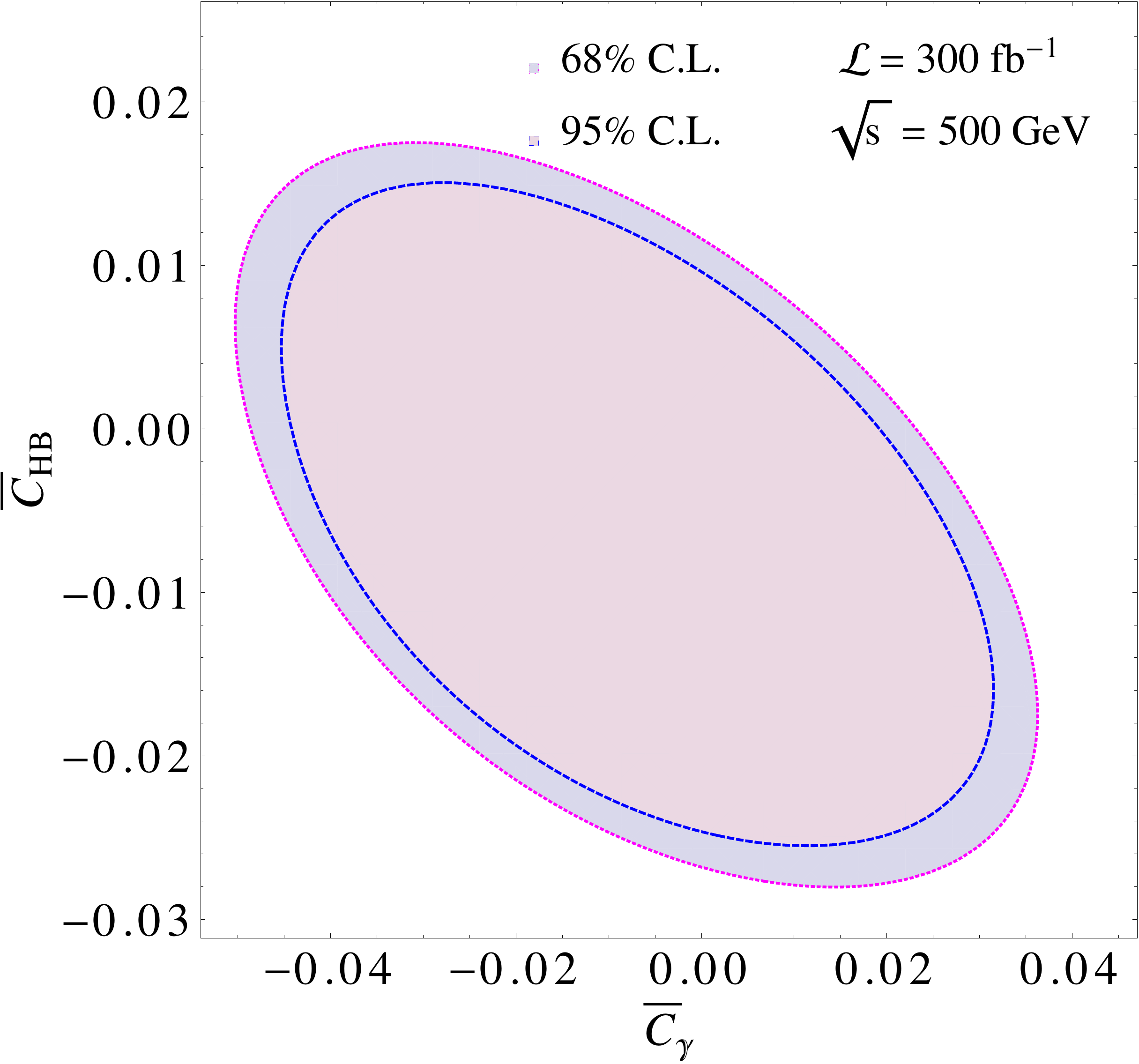}}   
		\resizebox{0.24\textwidth}{!}{\includegraphics{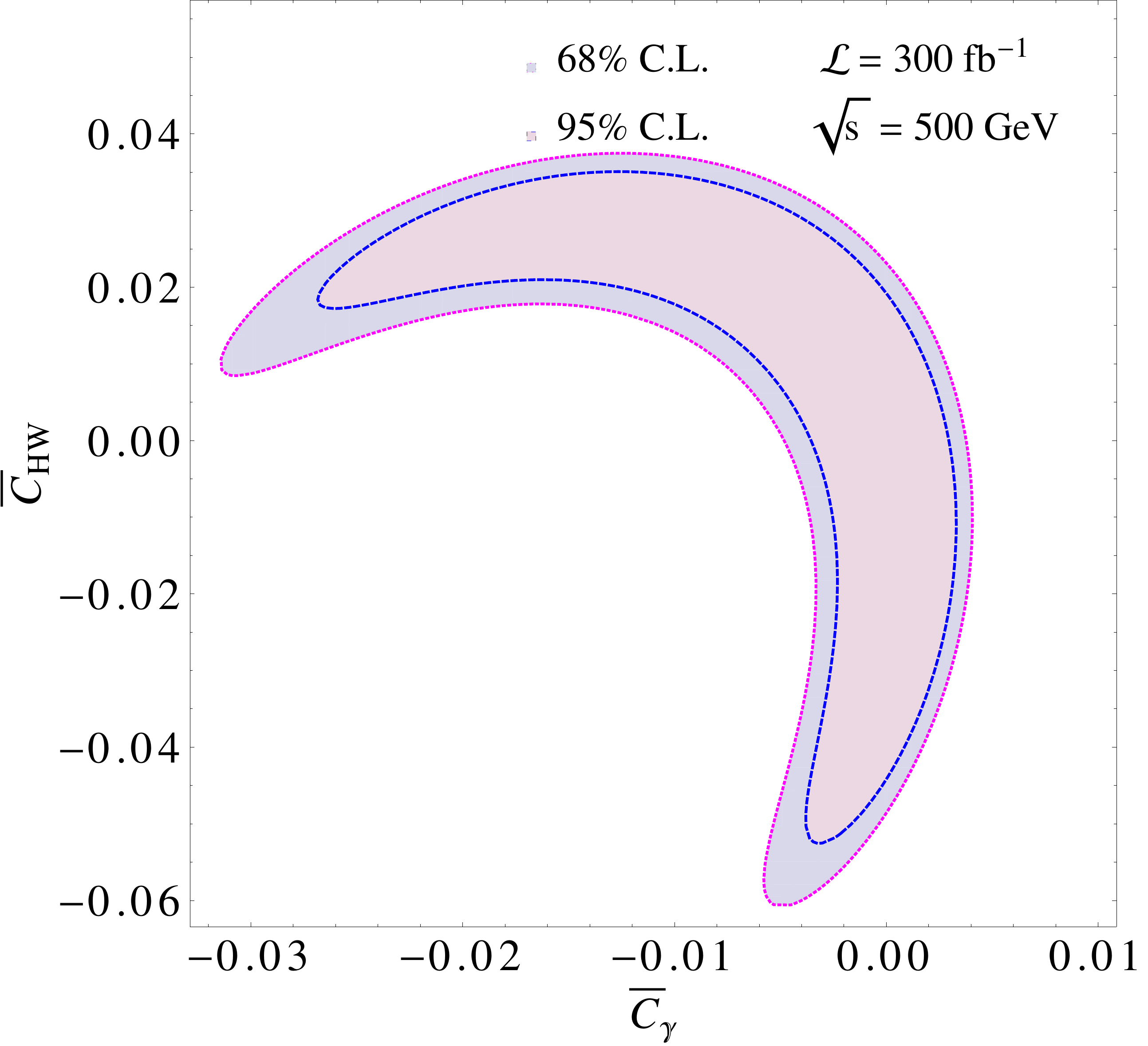}}   
		\resizebox{0.24\textwidth}{!}{\includegraphics{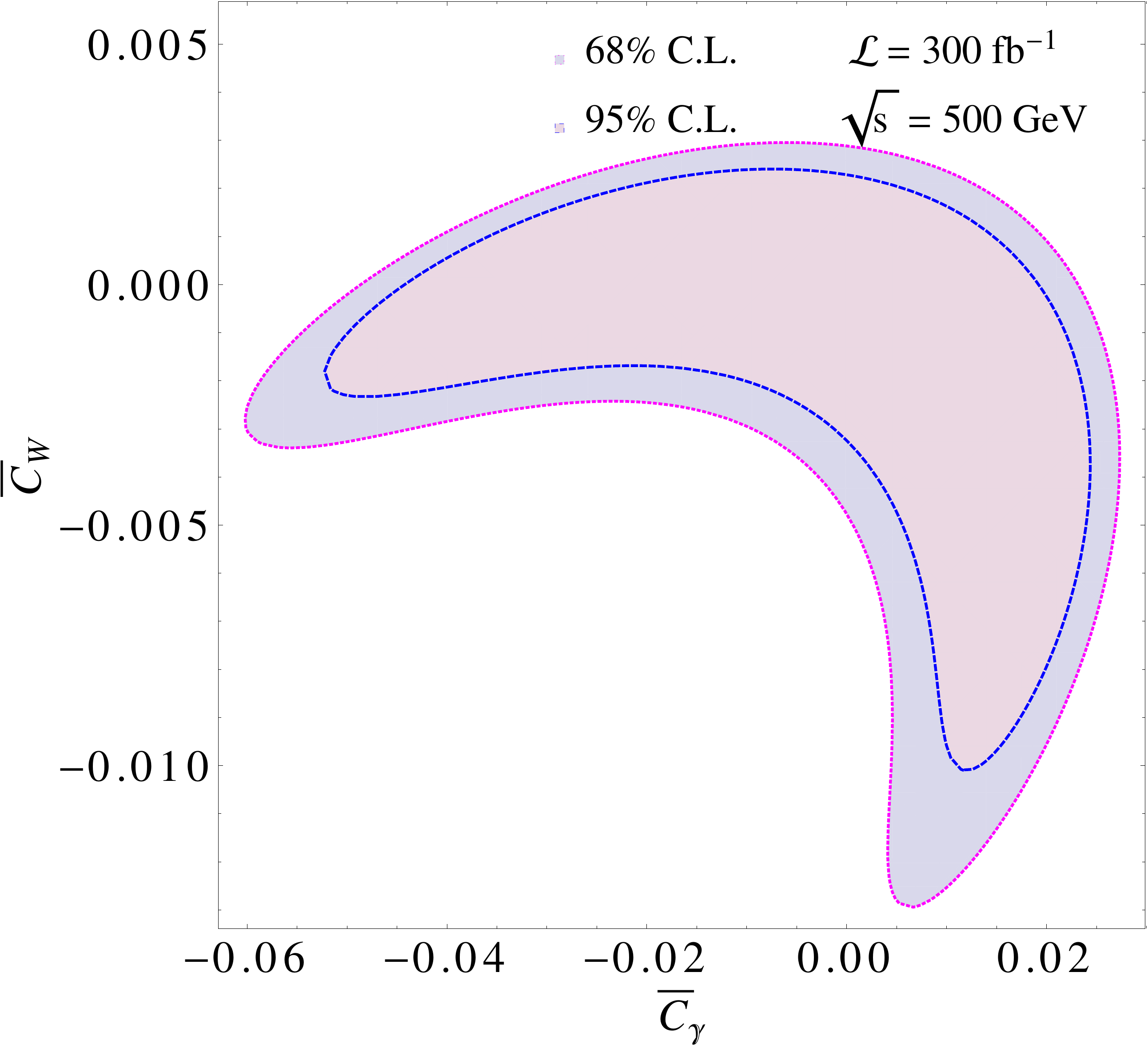}}   
		\vspace{0.5cm}		
		\resizebox{0.24\textwidth}{!}{\includegraphics{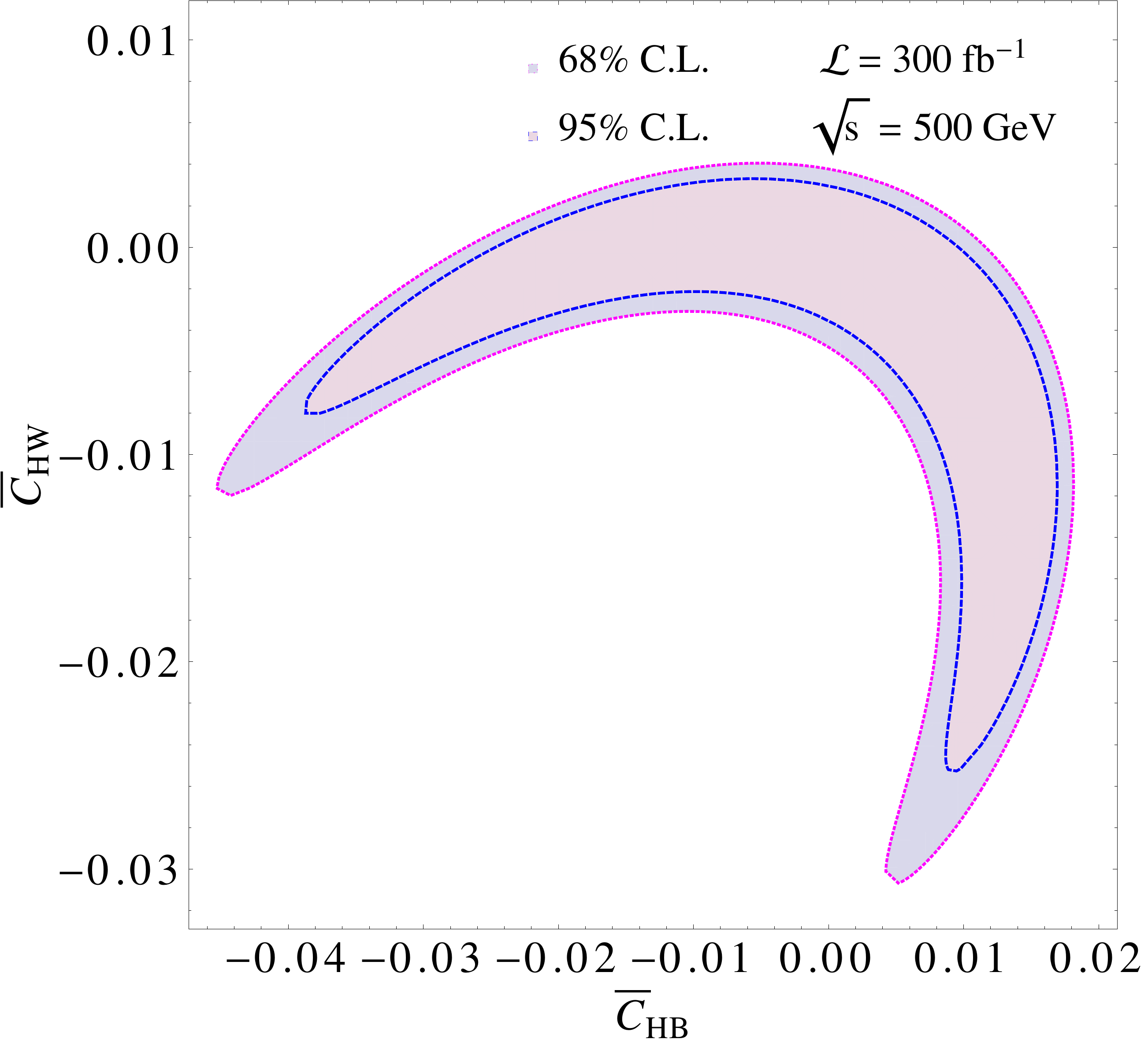}}   
		\resizebox{0.24\textwidth}{!}{\includegraphics{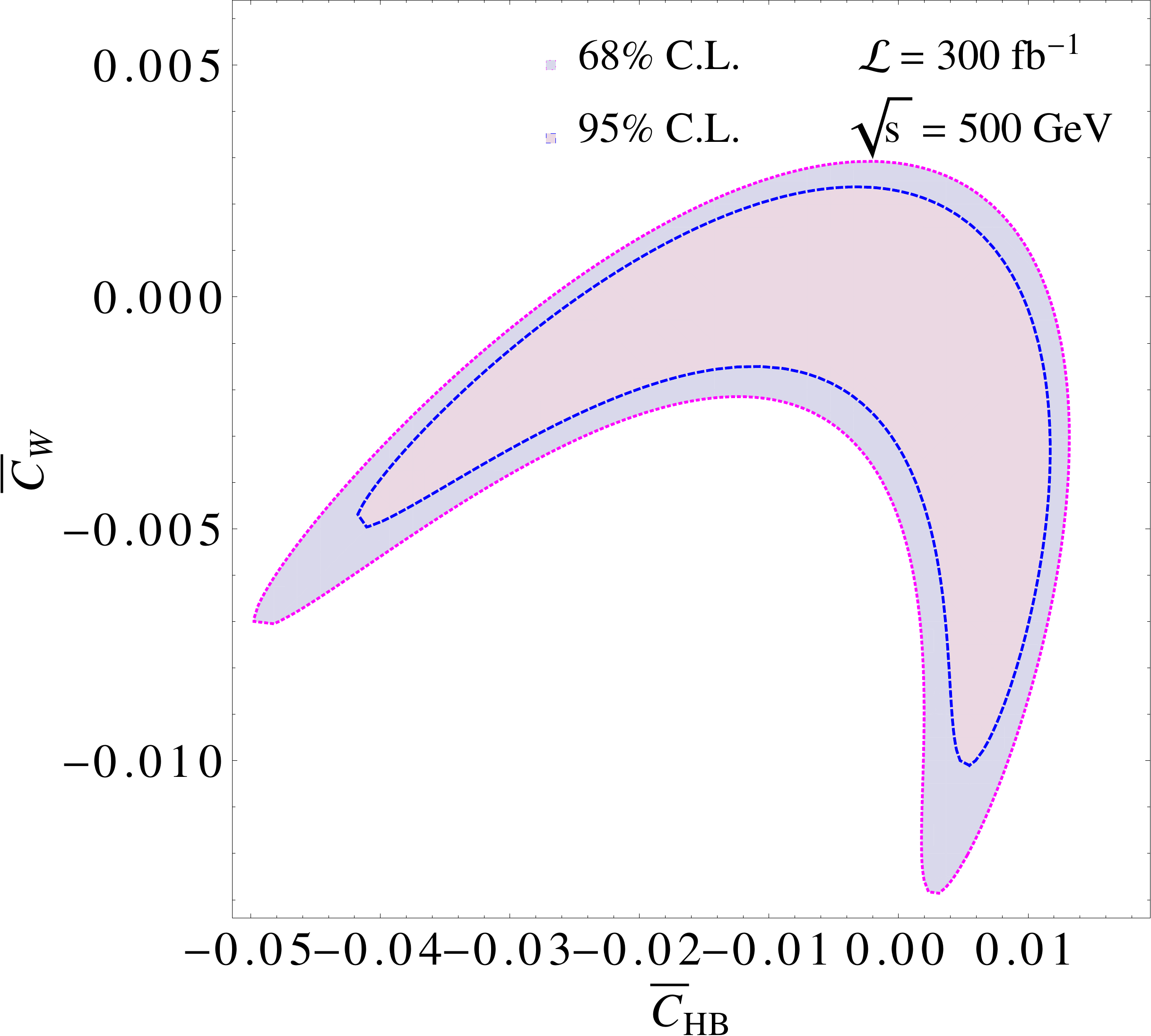}}   
		\vspace{0.5cm}		
		\resizebox{0.24\textwidth}{!}{\includegraphics{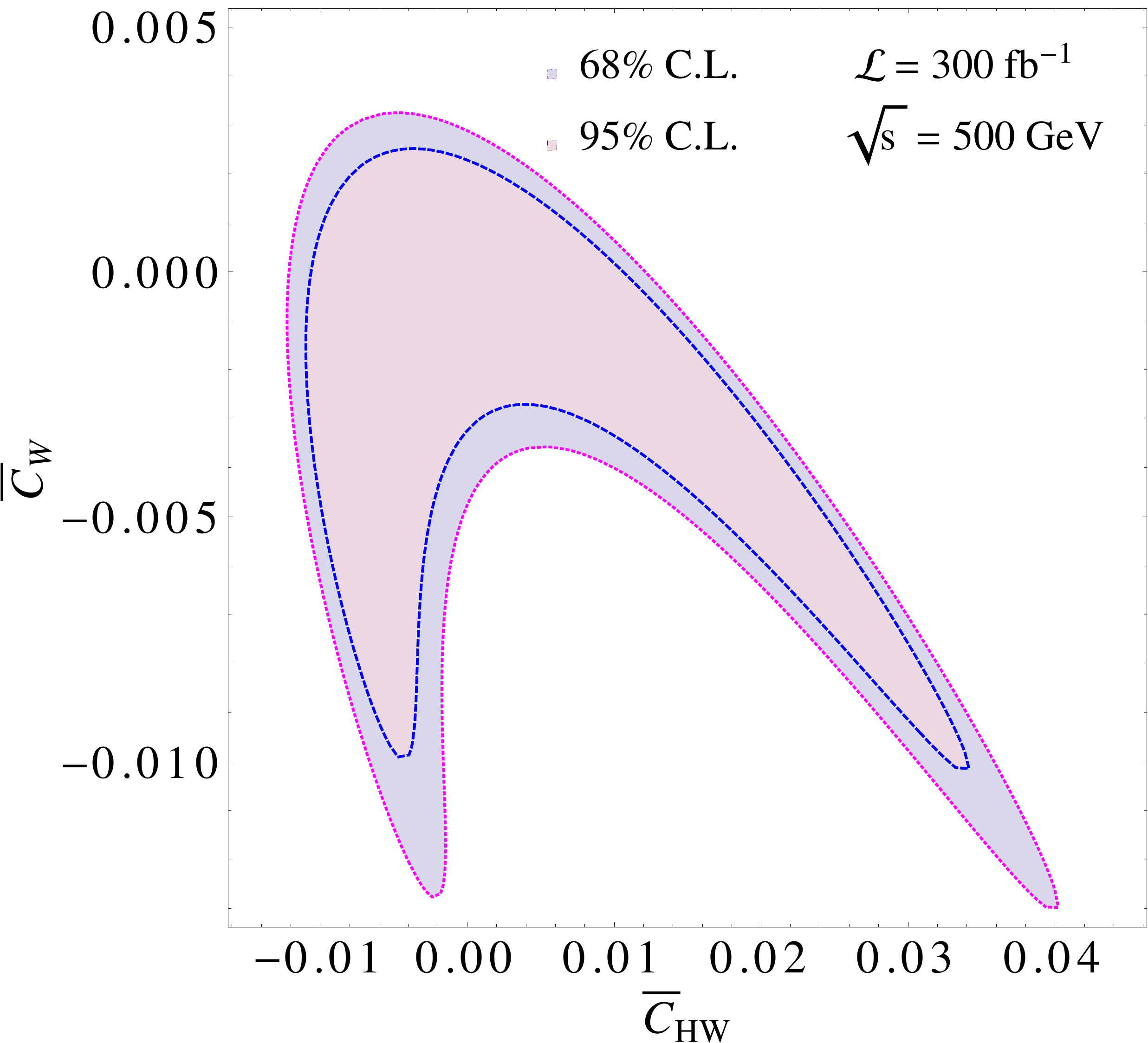}}   
		\caption{  Contours of 68\% and 95\% confidence level obtained from a fit using the
			$\cos (\ell, b-jets)$ distributions for the center-of-mass energy of $\sqrt{s} = 500$ GeV with an integrated luminosity of 300 fb$^{-1}$.  }\label{fig:X2-eeHZ-500GeV}
	\end{center}
\end{figure*}

In this section the results of the analysis are presented for the electron-positron collisions at the 
center-of-mass energies of 350 GeV and 500 GeV. 
The expected two dimensional contours at 68\% and 95\% confidence level on $(c_{i},c_{j})$ coefficients
are shown in Fig.~\ref{fig:X1-eeHZ-350GeV} and Fig.~\ref{fig:X2-eeHZ-350GeV} for the integrated luminosities 
of 300 fb$^{-1}$ and 3 ab$^{-1}$ of collisions at $\sqrt{s} = 350$ GeV. For comparison, the results are also presented 
for the center-of-mass energy collision of 500 GeV  in Fig.~\ref{fig:X2-eeHZ-500GeV}. 
For both integrated luminosities and both energies, one could clearly see that the limits on the coefficients ($\bar{c}_{ HW}$, $\bar{c}_{ W}$)
are considerably stronger than the other coefficients.  

At the center-of-mass energy of 350 GeV, for some Wilson coefficients, increasing the integrated luminosity from 300 fb$^{-1}$ to 3 ab$^{-1}$
can improve the constraints by a factor of around two and some by a factor of three. 

From Table~\ref{lhc}, where the bounds from one dimensional fit are extracted, 
it can be seen that going to higher energy of the electron-positron collisions, from 350 GeV to 500 GeV,  
would lead to improvements for the Wilson coefficients. For example, the constraints obtained from 350 GeV with the integrated luminosity of 
300 fb$^{-1}$ on $\bar{c}_{W}$ is $-0.00480 < \bar{c}_{W} < 0.00379$ which is tightened as $-0.00324 < \bar{c}_{W} < 0.00231$ at a 500 GeV machine.

It is instructive to compare the sensitivity of the bounds expected from high luminosity LHC with the bounds 
obtained here in this study.
In Table~\ref{lhc}, the results of this analysis are compared the ones expected to be achieved by the LHC
with the integrated luminosities of 300 fb$^{-1}$ and 3 ab$^{-1}$ for the case of  considering only one Wilson coefficient in the 
fit. In \cite{Englert:2015hrx}, the constraints  
on the Wilson coefficients have been obtained at the LHC at 14 TeV using the expected signal strength and the expected Higgs boson 
transverse momentum. The LHC bounds have been estimated using various Higgs boson production modes  and 
decay channels.
As it can be seen, while in this work only one Higgs production mode and decay has been considered,
more sensitivity is achievable on the coefficients $\bar{c}_{W}$ and $\bar{c}_{HW}$
at the electron-positron colliders with respect to the LHC. 
In this work,  the sensitivity to the dimension-six coefficients is obtained 
by considering only the Higgs boson decay into a $b\bar{b}$ pair. Including the 
other Higgs boson decay channels such as $H\rightarrow \gamma\gamma$,  
$H\rightarrow WW^{*}$, $H\rightarrow ZZ^{*} $, and $H\rightarrow \tau\tau $ 
provides significantly improved sensitivity to dimension-six coefficients. The hadronic and 
invisible decays of the $Z$ boson as well as using the $WW$-fusion Higgs production channels
would be significantly useful to improve the exclusion ranges at the electron-positron colliders.

It is notable that the next-to-leading order corrections~\cite{Fleischer:1982af,Kniehl:1991hk,Driesen:1995ew} to the  production cross section of $H+Z$ production 
could modify the shape of the $\cos(\ell, b)$ which needs to be considered in obtaining the sensitivity. To consider 
such effects, an overall large uncertainty of $10\%$ in each bin of  $\cos(\ell, b)$ distribution is taken into account
and the bounds are computed again. For example, the constraints on  $\bar{c}_{W}$ and $\bar{c}_{HW}$ at the 
center-of-mass energy of 350 GeV with an integrated luminosity of 3000 fb$^{-1}$ are as follows: 
$-0.00144 < \bar{c}_{W} < 0.00133$ and $-0.00196 < \bar{c}_{HW} < 0.00189$. Therefore, including a $10\%$ conservative
uncertainty would not weaken the limits significantly.

\begin{table*}[]
	\centering
	\caption{The expected bounds at 95\% CL on the Wilson coefficients from the LHC~\cite{Englert:2015hrx} 
	at the center-of-mass energy of 14 TeV with 300 fb$^{-1}$
		and 3000 fb$^{-1}$ as well the limits obtained form the current analysis in the 
		electron-positron collisions at the center-of-mass energies 
		of 350 GeV and 500 GeV considering only one coefficient in the fit.}
	\label{lhc}
	\begin{tabular}{l||lllll}
		\hline
		& LHC-300                            & LHC-3000    & $e^{-}e^{+}-350-300$   & $e^{-}e^{+}-350-3000$  & $e^{-}e^{+}-500-300$  \\   \hline     \hline  
		$\bar{c}_{W}[\times 10^{3}]$                 & {[}-8.0, 8.0{]}                     & {[}-4.0, 4.0{]} &      {[}-4.80, 3.79{]}         &    {[}-1.37, 1.27{]}                             &{[}-3.24, 2.31{]}                                \\
		$\bar{c}_{H}[\times 10^{3}]$                 & {[}\textless -50, \textgreater 50{]}  & {[}-44, 35{]} &      {[}-118.43, 129.85{]}       &    {[}-39.40, 40.70{]}                            &{[}-117.58, 145.86{]}                                \\
		$\bar{c}_{HW}[\times 10^{3}]$                & {[}-7.0, 10.0{]}                    & {[}-4.0, 4.0{]} &      {[}-6.19, 5.52{]}         &    {[}-1.87, 1.80{]}                                &{[}-3.65, 3.03{]}                                \\
		$\bar{c}_{\gamma} [\times 10^{3}]$            & {[}-1.9, 2.2{]}                     & {[}-0.6, 0.7{]} &      {[}-61.09, 19.78{]}       &    {[}-19.09, 6.25{]}                            &{[}-43.09, 19.64{]}                                \\
		$\bar{c}_{HB}[\times 10^{3}]$                & {[}-8.0, 11.0{]}                    & {[}-4.0, 4.0{]} &      {[}-51.35, 19.51{]}       &    {[}-17.20, 6.61{]}                              &{[}-24.70, 9.96{]}                                \\ \hline \hline 
	\end{tabular}
\end{table*}

The above bounds can be used to constrain the parameters of few  explicit 
models beyond the SM which at low energy limit reduce to the  
effective Lagrangian introduced in Eq.~\ref{eq:L}. In theories with strongly interacting Higgs boson, 
the Wilson coefficients are at the order of \cite{Giudice:2007fh,Contino:2013kra}:
\begin{eqnarray}
&& \bar{c}_{W} \sim O\Big(\frac{m_{W}}{M} \Big)^{2}, \bar{c}_{H}\sim O\Big(\frac{g^{\star}v}{M} \Big)^{2}, \nonumber \\ 
&& \bar{c}_{\gamma} \sim O\Big(\frac{m_{W}}{4\pi}\Big)^{2}\times \Big(\frac{y_{t}}{M}\Big)^{2}, \bar{c}_{HW} \sim O\Big(\frac{m_{W}}{4\pi}\Big)^{2}\times\Big(\frac{g^{\star}}{M}\Big)^{2}. \nonumber \\ 
\end{eqnarray}
where the strength of the Higgs boson coupling to a new physics state is denoted by $g^{\star}$ 
and $M$ is an overall mass scale of the new possible physical state at which the effective Lagrangian
is expected to be matched with the explicit models. As an example, translation of our constraint on  $\bar{c}_{W}$ leads to a lower limit of 2.3 TeV on the scale $M$.

%
\section{Summary and conclusions} \label{sec:Discussion}
%

Hints for  physics beyond the SM are expected to be found in  the Higgs  boson sector which in general
could lead to deviations in the Higgs boson couplings with respect to the SM predictions. 
As a result, indirect searches for new physics via Higgs boson require  precise measurement of  
the Higgs boson properties which could be performed by  future electron-positron colliders.

At the electron-positron colliders with the center-of-mass energy above the $m_{Z}+m_{H}$
threshold, large number of Higgs bosons could be  produced in association with $Z$ bosons. 
With the clean environment in the $e^{-}e^{+}$ colliders, the $H+Z$ events could be tagged 
easily through the leptonic $Z$  decays and Higgs bosons decays into $b\bar{b}$ pairs.
The expected very good resolution for leptons and jets momenta
measurements and identifications,  provides the possibility to characterize this final state
efficiently.  Therefore, a very precise measurement of the total and differential cross section of $H+Z$  can be performed at the future
electron-positron colliders. In this work, by performing a comprehensive analysis including the main sources 
of background processes and response of the detector, we find the potential of a future electron-positron 
collider to search for new physics originating from a complete set of effective dimension six operators that 
can contribute to Higgs boson production associated with a $Z$ boson.
We perform an  analysis on the differential cross section of the cosine of the angle between the most energetic
charged lepton from $Z$ boson decay and the most energetic  b-jet from the Higgs boson decay 
to find the sensitivity of $e^{-}+e^{+} \rightarrow H+Z$ process to the  
dimension six operators. The analysis is done 
at the center-of-mass energies of 350 GeV and 500 GeV with an ILD-like detector considering the integrated 
luminosities of 300 fb$^{-1}$ and 3 ab$^{-1}$. 
It is found that the $e^{-}+e^{+} \rightarrow H+Z$ process has a great sensitivity 
to dimension six  operators induced at tree level.  We show that 
high luminosity runs of the future electron-positron colliders would be able to improve the sensitivity of high luminosity LHC 
to new physics via Higgs boson.

%
\section*{Acknowledgments}
%

The authors are especially grateful to Sara Khatibi for the fruitful discussions. Authors thank School of Particles and Accelerators, 
Institute for Research in Fundamental Sciences (IPM) for financial support of this project.
Hamzeh Khanpour also is thankful the University of Science and Technology of Mazandaran for financial support provided for this research.

\appendix

%
%
\section{Cut flow table for the center-of-mass energy of 350 GeV}\label{AppendixA}


Table~\ref{Table:Cut-Table2} presents the expected cross sections 
after different combinations of cuts for signal and SM background processes. The numbers are given 
in the unit of fb. The signal cross sections are corresponding to particular values 
of $\bar{c}_H = 0.1$ and  $\bar{c}_\gamma = 0.1$. The center-of-mass energy of the collision is assumed to be $\sqrt s = 350$ GeV.

\begin{table*}[htbp]
	\begin{center}
		\begin{tabular}{c|c c|c c c c}
			$\sqrt s = 350$ GeV      &  \multicolumn{2}{c|}{Signal }    & \multicolumn{2}{c}{~~~~~~~~~~~~~~~~~~~~~~~ Background }    \\   \hline
			Cuts    & $\bar{c}_H$    & $\bar{c}_\gamma$          & SM ($H+Z$) & $t \bar{t}$ &  $ZZ$   &  $Z\gamma,\gamma\gamma,WWZ$   \\ \hline  \hline
			Cross-sections (in fb)              &  $10.21$  &  $26.46$  & $11.30$        & $10.42$   &  $59.42$     &  $20.62$     \\
			(I):	2$\ell$, $|\eta^{\ell}|<2.5$, $p_T^{\ell}>10$    &  $7.33$  &  $18.63$ & $8.10$  &   $6.17$  &  $40.86$  &  $7.74$    \\
			(II):	$2{\rm jets}$, $|\eta^{\rm jet}|<2.5$, $p_T^{\rm jet}>20$, $\Delta R_{\ell, \rm jet}\geq 0.5$   &  $5.09$ &  $12.99$   & $5.61$   &  $4.58$   &  $24.44$      &  $5.76$      \\
			(III): $2{\rm b-jets}$  &  $2.00$ &  $5.10$   & $2.21$   &  $1.73$   &  $1.87$      &  $2.30$      \\ \hline 			
			(IV): $p_T^{\ell^+ \ell^-}>100$    &  $1.55$  &  $3.64$ & $1.71$  &   $0.08$  &  $0.81$  &  $0.46$    \\
			(V): $90 <  m_{b \bar{b}} < 160,~75 <  m_{\ell^+ \ell^-} < 105$ &  $1.284$   &  $2.997$   &  $1.410$     &  $0.003$ &   $0.122$       &  $0.016$     \\          \hline         \hline
		\end{tabular}
	\end{center}
	\caption{ Expected cross sections in unit of fb after different combinations of cuts for signal and SM background processes. 
		The signal cross sections are corresponding to particular values of $\bar{c}_H = 0.1$ and  $\bar{c}_\gamma = 0.1$. 
		The center-of-mass energy of the collision is assumed to be 350 GeV. More details of the selection cuts are given in
		 Section \ref{sec:analysis}. }
	\label{Table:Cut-Table2}
\end{table*}

%
%


%

\end{document}